\documentclass[sigconf,nonacm,pdfa]{acmart}

\newcommand\vldbdoi{10.14778/3665844.3665846}
\newcommand\vldbpages{2136 - 2148}
\newcommand\vldbvolume{17}
\newcommand\vldbissue{9}
\newcommand\vldbyear{2024}
\newcommand\vldbauthors{\authors}
\newcommand\vldbtitle{\shorttitle} 
\newcommand\vldbavailabilityurl{https://spatialyze.github.io}
\newcommand\vldbpagestyle{empty} 

\usepackage{CJKutf8}
\usepackage{algorithm}
\usepackage{xspace}
\usepackage{soul}
\usepackage[export]{adjustbox}

\newcommand{\bm}[1]{#1}
\newcommand{\bmv}[1]{#1}

\newcommand{\inview}{Road Visibility Pruner\xspace}
\newcommand{\objtype}{Object Type Pruner\xspace}
\newcommand{\depthEstimation}{Geometry-Based 3D Location Estimator\xspace}
\newcommand{\sample}{Exit Frame Sampler\xspace}

\newcommand{\tcc}[1]{#1\xspace}
\newcommand{\tcs}[1]{#1\xspace}

\newcommand{\cWorld}{\tcc{World}}

\newcommand{\cGeoConstruct}{\tcc{Geographic Construct}}
\newcommand{\cGeoConstructs}{\tcc{Geographic Constructs}}
\newcommand{\cObject}{\tcc{Movable Object}}
\newcommand{\cObjects}{\tcc{Movable Objects}}

\newcommand{\sWorld}{\tcs{World}}

\newcommand{\sCamera}{\tcs{Camera}}
\newcommand{\sCameras}{\tcs{Cameras}}
\newcommand{\sVideo}{\tcs{Video}}
\newcommand{\sVideos}{\tcs{Videos}}
\newcommand{\sGSVideo}{\tcs{GeospatialVideo}}

\newcommand{\sRoadNetwork}{\tcs{RoadNetwork}}
\newcommand{\sRoadNetworks}{\tcs{RoadNetworks}}

\newcommand{\cstar}[1]{{\LARGE\color{#1}$\star$}\color{black}\xspace}
\newcommand{\circled}[1]{{\textcircled{\footnotesize{#1}}}}
\newcommand{\bigcircled}[1]{{\textcircled{\small{#1}}}}

\newcommand{\newvspace}[1]{\vspace{#1}}

\newcommand{\eqspaceValue}{2pt}
\newcommand{\eqspace}{\setlength{\belowdisplayskip}{\eqspaceValue} \setlength{\belowdisplayshortskip}{\eqspaceValue}\setlength{\abovedisplayskip}{\eqspaceValue} \setlength{\abovedisplayshortskip}{\eqspaceValue}}

\newcommand{\aboveCodeVSpace}{\newvspace{-8pt}}
\newcommand{\belowCodeVSpace}{\newvspace{-8pt}}

\newcommand{\aboveCapVSpaceTwo}{}
\newcommand{\belowCapVSpace}{}
\newcommand{\tableAboveVSpace}{}
\newcommand{\tableBelowVSpace}{}

\renewcommand{\aboveCapVSpaceTwo}{\newvspace{-5pt}}
\renewcommand{\tableAboveVSpace}{\newvspace{-6pt}}

\renewcommand{\sectionautorefname}{\S\kern-1.5pt}
\renewcommand{\subsectionautorefname}{\S\kern-1.5pt}
\renewcommand{\subsubsectionautorefname}{\S\kern-1.5pt}

\newcommand{\NumVideos}{240}
\newcommand{\NumScenes}{80}

\newcommand{\evalVariableZeroOverallRuntime}{34}

\newcommand{\evalVariableTwoIngestPercent}{0.01}

\newcommand{\evalVariableOneQueryPercent}{9.5}

\newcommand{\evalVariableOneSavePercent}{0.6}

\newcommand{\evalVariableZeroVProcessPercent}{90}
\newcommand{\evalVariableOneVProcessPercent}{89.9}

\newcommand{\evalVariableOneInviewPercent}{0.1}

\newcommand{\evalVariableOneInviewQueryThreeFourSkip}{3.8}

\newcommand{\evalVariableOneInviewQueryThreeFourRuntimeReduction}{3.2}

\newcommand{\evalVariableOneInviewQueryOneSkip}{21.5}

\newcommand{\evalVariableOneInviewQueryTwoRuntimeReduction}{19.6}

\newcommand{\evalVariableOneInviewQueryOneTwoRuntimeReduction}{20.0}

\newcommand{\evalVariableZeroObjTypeQueryOneSSRuntimeReduction}{69}

\newcommand{\evalVariableZeroObjTypeQueryOneRuntimeReduction}{26}
\newcommand{\evalVariableOneObjTypeQueryOneRuntimeReduction}{26.3}

\newcommand{\evalVariableOneObjTypeQueryOneObjReduction}{86.3}

\newcommand{\evalVariableTwoObjTypeQueryTwoRuntimePercent}{0.06}

\newcommand{\evalVariableOneObjTypeQueryTwoObjReduction}{36.5}

\newcommand{\evalVariableOneObjTypeQueryTwoThreeFourRuntimeReduction}{7.0}

\newcommand{\evalVariableZeroMonodepthAllPercent}{48}

\newcommand{\evalVariableTwoDepthAllTDEstmReductionPercent}{99.48}

\newcommand{\evalVariableTwoDepthAllTDEstmPortionPercent}{0.55}
\newcommand{\evalVariableZeroDepthAllTDEstmSpeedup}{192}

\newcommand{\evalVariableZeroDepthAllRuntimeReduction}{52}

\newcommand{\evalVariableOneDeQueryTwoSkipPercent}{13.1}

\newcommand{\evalVariableOneDeAllRuntimeReductionPercent}{0.8}

\newcommand{\evalVariableOneOverallQueryThreeFourRuntimeSpeedup}{2.5}

\newcommand{\evalVariableOneOverallQueryOneRuntimeSpeedup}{5.3}

\newcommand{\evalVariableTwoOverallDeQueryTwoAdditionalSpeedup}{1.65}

\newcommand{\evalVariableTwoOverallDeQueryThreeFourAdditionalSpeedup}{4.32}

\newcommand{\evalVariableReductionTwentyDistance}{28.27}
\newcommand{\evalVariableOneDeQueryTwoSkipDistance}{3.6}
\newcommand{\evalVariableOneDeAllSSReductionDistance}{39}

\newcommand{\evalVariableOneQueryOneInViewAccuracyDrop}{4.7}

\newcommand{\evalVariableOneQueryOneObjectFilterAccuracyDrop}{5.3}

\newcommand{\evalVariableOneQueryTwoObjectFilterAccuracyDrop}{2.5}

\newcommand{\evalVariableOneQueryTwoOptimizedwithDetectionEstimationAccuracy}{83.4}

\newcommand{\evalVariableOneQueryThreeFourInViewAccuracyDrop}{0.4}

\newcommand{\evalVariableOneQueryThreeFourOptimizedAccuracy}{97.1}

\newcommand{\evalVariableOneAllOptimizedAccuracy}{93.4}

\newcommand{\evalVariableOneAllOptimizedDEAccuracy}{84.5}

\newcommand{\UseTitle}[1]{\newcommand{\PaperTitle}{#1}}
\UseTitle{Spatialyze: A Geospatial Video Analytics System with~Spatial-Aware Optimizations}

\usepackage{listings}
\definecolor{hlgray}{rgb}{0.832,0.832,0.832}
\renewcommand{\circled}[1]{(#1)}
\renewcommand{\bigcircled}[1]{\sethlcolor{hlgray}\hl{(#1)}}
\renewcommand{\Description}[1]{}
\definecolor{codelightgreen}{rgb}{0.29,0.47,0.48}
\definecolor{codegray}{rgb}{0.5,0.5,0.5}
\definecolor{codered}{rgb}{0.67,0.19,0.16}
\definecolor{codegreen}{rgb}{0.21,0.49,0.13}
\definecolor{codeblue}{rgb}{0,0.02,0.96}
\lstdefinestyle{mystyle}{
    language=Python,
    commentstyle=\color{codelightgreen},
    keywordstyle=\bfseries\color{codegreen},
    numberstyle=\tiny\color{codegray},
    stringstyle=\color{codered},
    basicstyle=\ttfamily\scriptsize,
    breakatwhitespace=false,
    keepspaces=true,
    numbers=left,
    numbersep=5pt,
    showspaces=false,
    showstringspaces=false,
    showtabs=false,
    basewidth=0.5em,
    emph={filter,object,type},
    emphstyle=\ttfamily,
    emph={[2]exitsLane,exitsCamer,newCar},
    emphstyle={[2]\color{codeblue}},
    frame=tb,
    framesep=0pt,
    framextopmargin=2pt,
    tabsize=2
}
\renewcommand{\aboveCodeVSpace}{\newvspace{-3pt}}
\renewcommand{\belowCodeVSpace}{\newvspace{-3pt}}

\begin{document}
\title{\PaperTitle}

\newcommand{\ucberkeley}{UC Berkeley}

\settopmatter{authorsperrow=4}
\author{\mbox{Chanwut Kittivorawong}}
\authornote{Co-first authors.}
\affiliation{\institution{\ucberkeley}}
\email{chanwutk@berkeley.edu}

\author{Yongming Ge}
\authornotemark[1]
\affiliation{\institution{\ucberkeley}}
\email{yongmg@berkeley.edu}

\author{Yousef Helal}
\affiliation{\institution{\ucberkeley}}
\email{yousefh@berkeley.edu}

\author{Alvin Cheung}
\affiliation{\institution{\ucberkeley}}
\email{akcheung@cs.berkeley.edu}

\begin{abstract}
Videos that are shot using commodity hardware such as phones and surveillance cameras record various metadata such as time and location.
We encounter such {\em geospatial videos} on a daily basis and such videos have been growing in volume significantly.
Yet, we do not have data management systems that allow users to interact with such data effectively.

In this paper, we describe Spatialyze, a new framework for end-to-end querying of geospatial videos.
Spatialyze comes with a domain-specific language where users can construct geospatial video analytic workflows using a 3-step, declarative, {\em build-filter-observe} paradigm.
Internally, Spatialyze leverages the declarative nature of such workflows, the temporal-spatial metadata stored with videos, and physical behavior of real-world objects to optimize the execution of workflows.
Our results using real-world videos and workflows show that Spatialyze can reduce execution time by up to \bmv{\evalVariableOneOverallQueryOneRuntimeSpeedup$\times$}, while maintaining up to 
\bmv{\evalVariableOneQueryThreeFourOptimizedAccuracy\%}
accuracy compared to unoptimized execution.

\end{abstract}

\maketitle
\pagestyle{\vldbpagestyle}
\begingroup\small\noindent\raggedright\textbf{PVLDB Reference Format:}\\
\vldbauthors. \vldbtitle. PVLDB, \vldbvolume(\vldbissue): \vldbpages, \vldbyear.\\
\href{https://doi.org/\vldbdoi}{doi:\vldbdoi}
\endgroup
\begingroup
\renewcommand\thefootnote{}\footnote{\noindent
This work is licensed under the Creative Commons BY-NC-ND 4.0 International License. Visit \url{https://creativecommons.org/licenses/by-nc-nd/4.0/} to view a copy of this license. For any use beyond those covered by this license, obtain permission by emailing \href{mailto:info@vldb.org}{info@vldb.org}. Copyright is held by the owner/author(s). Publication rights licensed to the VLDB Endowment. \\
\raggedright Proceedings of the VLDB Endowment, Vol. \vldbvolume, No. \vldbissue\ %
ISSN 2150-8097. \\
\href{https://doi.org/\vldbdoi}{doi:\vldbdoi} \\
}\addtocounter{footnote}{-1}\endgroup

\ifdefempty{\vldbavailabilityurl}{}{
\vspace{.3cm}
\begingroup\small\noindent\raggedright\textbf{PVLDB Artifact Availability:}\\
The source code, data, and/or other artifacts have been made available at \url{\vldbavailabilityurl}.
\endgroup
}

\section{Introduction}

Geospatial videos record the locations and periods of the recordings in addition to the visual information.
Such videos are prevalent in our daily lives,
from surveillance footage to autonomous vehicle (AV) cameras to police body cameras.
While the volume of such data has grown tremendously~\cite{wright:av-data, sparks:security-cam},
we still lack end-to-end systems that can process and query such data effectively.
The rise of machine learning (ML) in recent years has aggravated this issue,
with the latest deep learning models capable of carrying out various computer vision tasks, 
such as object detection~\cite{girshick:rcnn, josher:yolov5, girshick:fast-rcnn, redmon:yolov3, ren:faster-rcnn},
multi-objects tracking~\cite{bewley:sort, du:strongsort, wojke:deepsort, brostrom:yolo-strongsort},
image depth estimation~\cite{godard:monodepth}, etc.

All these trends have made geospatial video analytics computationally intensive. For instance, 
a \$2,299
NVIDIA T4 GPU takes \bmv{\evalVariableZeroOverallRuntime} seconds on average to execute a simple workflow of object detection, tracking, and image depth estimation on a
20-second 12-fps video.
Modern autonomous driving datasets, such as nuScenes~\cite{caesar:nuscenes}, contain 6000 such videos.
Running the workflow mentioned above will take 3 full days to run on the entire dataset.

To make matters worse, the lack of programming frameworks and data management systems for geospatial videos has made it challenging for end users to specify their workflows, let alone run them efficiently.
For instance, a data journalist writing an article on self-driving car failures 
would like to examine the footage collected on car cameras to look for a specific behavior, e.g., two cars crossing at an intersection.
They will either need to watch all the footage themself or string together various ML models for video analytics~\cite{kang:noscope,bastani:miris,bastani:otif,xu:eva}
and video-processing libraries
(e.g., OpenCV~\cite{bradski:opencv}, FFmpeg~\cite{suramya:ffmpeg})
to construct their workflow.
Sadly, the former is infeasible given the amount of video data collected,
while the latter requires programming expertise that most users do not possess.

We aim to build a programming framework that bridges video processing and geospatial video analytics.
Specifically, as users constructing workflows on geospatial videos are typically interested in the time and location where such videos are taken,
and such information is often stored as video metadata, we exploit that to optimize such workflows.
Objects captured in videos inherit the physical behavior of real-world objects.
For example, when users look for a car at a road intersection,
the intersection must be visible in the video frames that the car of interest is visible;
therefore, video frames without a visible intersection,
which can be determined using geospatial metadata,
can be skipped during query processing.
We leverage such existing geospatial metadata, the {\em inherited physical behaviors},
and users' queries to speed up video processing.

Leveraging this insight, we present Spatialyze, a system for geospatial video analytics.
To make it easy for users to specify their geospatial video analytics workflows,
Spatialyze comes with a conceptual data model where users create and ingest videos into a ``world,''
and users interact with the world by specifying objects (e.g., cars)
and scenarios (e.g., cars at an intersection) of interest via Spatialyze's S-Flow,
a domain-specific language (DSL) embedded in Python. 
Spatialyze then efficiently executes the workflow by leveraging various spatial-aware optimization techniques that make use of existing geospatial metadata and assumptions based on the {inherited physical behavior}
of objects in the videos.
For instance, Spatialyze's {\em \inview} uses the road's visibility as a proxy for objects' visibility to prune out video frames.
The {\em \objtype} then prunes out objects that are not of interest to the users' workflow.
Spatialyze's {\em \depthEstimation} speeds up object 3D location estimation by replacing a computationally expensive ML-based approach with a geometry-based approach.
Finally, the {\em \sample} prunes out unnecessary video frames based on the {inherited physical behavior} of vehicles and traffic rules.
Spatialyze's optimizations are all driven by various geospatial metadata and real-world physical behavior,
with the goal of reducing the number of video frames to be processed and ML operations to be invoked,
thus speeding up the video processing runtime.

As we are unaware of end-to-end geospatial video analytics systems,
we evaluate different parts of Spatialyze against state-of-the-art (SOTA) video analytic tools (OTIF~\cite{bastani:otif}, VIVA~\cite{romero:viva}, EVA~\cite{xu:eva}),
a geospatial data analytic tool (nuScenes devkit~\cite{caesar:nuscenes}), and an aerial drone video sensing platform (SkyQuery~\cite{bastani:skyquery}) on different geospatial video analytics workflows.
We are up to \bmv{7.3$\times$} faster than EVA in geospatial object detection workflow,
\bmv{1.06-2.28$\times$} faster than OTIF in object tracking,
\bmv{1.68$\times$} faster than VIVA,
\bmv{1.18$\times$} faster than SkyQuery in geospatial object tracking workflow,
and \bm{117-716$\times$} faster than nuScenes in geospatial data analytics.
We also evaluate each of our optimization techniques by executing each of the queries with and without each
of the techniques,
shown in the results,
achieving
\bmv{\evalVariableOneOverallQueryThreeFourRuntimeSpeedup-\evalVariableOneOverallQueryOneRuntimeSpeedup$\times$} speed up with 
\bmv{\evalVariableOneQueryTwoOptimizedwithDetectionEstimationAccuracy-\evalVariableOneQueryThreeFourOptimizedAccuracy\%}
accuracy on 3D object tracking.

In sum, we make the following contributions.
{In \autoref{sec:ql}},
we present a conceptual data model and a DSL (S-Flow) for geospatial video analytics.
Using S-Flow, users construct their analytic workflows simply by following our \emph{build}-\emph{filter}-\emph{observe} paradigm.
{In \autoref{sec:execution}},
we describe the design of Spatialyze,
a fast and extensible system to process geospatial videos of arbitrary length in S-Flow.
{In \autoref{sec:optimization}},
we develop 4 optimization techniques that leverage geospatial metadata embedded in videos and {inherited physical behavior} of objects to speed up geospatial video processing in Spatialyze's geospatial video analytic workflow.
{In \autoref{sec:evaluation}},
we implemented our techniques in Spatialyze and evaluated it against other SOTA video processing techniques, as well as ablation studies,
using real-world videos from nuScenes~\cite{caesar:nuscenes}, VIVA~\cite{romero:viva}, and SkyQuery~\cite{bastani:skyquery}.

\section{Related Work} \label{sec:related}

\paragraph{Geospatial Video Analytics}
Recent use cases for querying geospatial video data
(autonomous driving~\cite{kim:scenic-val},
surveillance~\cite{griffin:amber},
traffic analysis~\cite{gloudemans:vehicle-turn-count},
and transshipment activities~\cite{mcdonald:satellites-fishing, park:satellite-fishing})
have led to the development of geospatial DBMSs for data storage and retrieval.
VisualWorldDB~\cite{haynes:visualworlddb} allows users to ingest video data from various sources and make them queryable as a single multidimensional object.
It optimizes query execution by jointly compressing overlapping videos and reusing results from them.
Apperception~\cite{ge:apperception} provides a Python API for users to input geospatial videos, organize them into a four-dimensional data model, and query video data.
While Apperception focuses on organizing and retrieving geospatial video data,
it falls short of optimizing query execution.
In contrast, Spatialyze provides a comprehensive geospatial video analytic workflow interface for declaratively specifying videos of interest.
In addition, it introduces new query optimization techniques by leveraging the inherent geospatial properties of videos.

\paragraph{Video Processing}
Researchers have proposed techniques to efficiently detect, track, and estimate 3D locations of objects from videos.
Monodepth2~\cite{godard:monodepth} estimates per-pixel image depth data using only monocular videos.
EVA~\cite{xu:eva} identifies, materializes, and reuses expensive user-defined functions (UDF) on videos for exploratory video analytics with multiple refined and reused queries.
However, EVA's optimization techniques aim at scenarios where multiple queries repeatedly invoke overlapping UDFs.
OTIF~\cite{bastani:otif} is a general-purpose object tracking algorithm.
It uses a segmentation proxy model to determine if frames or regions of frames contain objects that need to be processed with an object detector,
along with a recurrent reduced-rate tracking method that speeds up tracking by reducing frame rate.
VIVA~\cite{kang:viva,romero:viva} is a video analytic system that allows users to specify
relationship opportunities of replacing or filtering one ML model with another.
It uses these relationships to speed up the video processing time.
However, it does not optimize video processing with geospatial metadata by default,
leaving substantial performance gains unexploited, as our experiments show.
Complexer-YOLO~\cite{simon:complexer-yolo} and
EagerMOT~\cite{kim:eagermot} propose one-step ML models for efficiently tracking 3D objects using LiDAR data,
which is larger and not always available compared to cameras and GPS sensors information.
Spatialyze overcomes this limitation by not requiring such information to process geospatial workflows,
but instead using readily available geospatial metadata to optimize user workflows.
In addition, one of our optimization techniques estimates objects' 3D location \bmv{\evalVariableZeroDepthAllTDEstmSpeedup$\times$} faster than the approach that uses Monodepth2 on average.
SkyQuery~\cite{bastani:skyquery} is a drone video sensing platform.
It also provides a tool for querying and visualizing sensed aerial drone videos but does not optimize video query processing.

\section{Usage Scenario}
\label{sec:example}

In this section, we give an overview of Spatialyze using an example.
Suppose a data journalist is interested in investigating AV footage reports,
and the goal is to find video snippets of other vehicles that crash into the driver's vehicle at a road intersection.

Naively, the data journalist would watch all the videos individually to
identify parts of the videos that are relevant.
Even with programming skills, they still need to manually invoke ML models to process large volumes of geospatial videos, and possibly optimize model inference by reducing the number of frames to be processed.
They need to then merge these results with geospatial metadata (e.g., road information) to extract vehicle geospatial data, write queries to isolate objects of interest, and link them to the geospatial objects for a cohesive output.
This process demands high coding expertise and is slow, tedious, and error-prone.

Spatialyze streamlines this process.
Instead of manually writing code to run ML models and extract geospatial information,
journalists can use Spatialyze to construct such workflows easily.
With Spatialyze, they can construct an analytic workflow with S-Flow's 3-step paradigm as shown in \autoref{listing:python-workflow} to:
\textbf{\emph{Build a world}}, by integrating the video data with the geospatial metadata
(e.g., road network and camera configurations);
\textbf{\emph{Filter the video for parts of interest}}, by describing relationships between objects
(humans, cars, trucks) and their surrounding geospatial environment (lane, intersection);
\textbf{\emph{Observe the filtered video parts}}, in this case,
by saving all the filtered video parts into video files for further examination.

{\em Build a world.}
In line~\ref{line:world}, the user creates a {\small \tt world} which represents a geospatial virtual environment to be discussed in \autoref{sec:ql}.
Line~\ref{line:road} then initializes the {\small \tt world} by loading {\small \tt roadNetwork} data,
in this case,
a set of the Boston Seaport's road segments.
Lines~\ref{line:gsvideo}-\ref{line:end-build} load videos with their camera configurations (locations, rotations, and intrinsics~\cite{zhang:camera}) and associate them to create geospatial videos added into the {\small \tt world}.
Finally, in line~\ref{line:end-build}, they have constructed a data-rich {\small\tt world} with %
video cameras, and road segments, all related spatially and temporally.

{\em Filter for the video parts of interest.}
From the data journalist's perspective, they can interact with objects and their trackings from each video added to the {\small \tt world} through {\small \tt object()} with the world initialized.
Here, our journalist wants to find video snippets where other vehicles crash into the driver's vehicle at a road intersection.
They do so by adding a filter to the {\small \tt world} to only include objects that match the provided predicates: that they
are cars or trucks (line~\ref{line:pred-type}),
are within 50 meters of the camera (line~\ref{line:pred-distance}),
are at an intersection (line~\ref{line:pred-contain}),
and are moving towards the camera (in line~\ref{line:pred-heading}).
After {\small\tt filter} is executed at line~\ref{line:pred-heading}, the {\small \tt world} will contain only cars or trucks on intersections that are moving towards the camera, as desired.

{\em Observe the filtered video parts.}
After applying the filter, the {\small \tt world} has fewer objects than when it was initialized.
When the journalist calls {\small \tt saveVideos()} in line~\ref{line:observe}, Spatialyze saves only video snippets with objects in the current {\small \tt world}, significantly shortening the total video length.
This allows the journalist to focus only on relevant segments with highlighted objects of interest.
As shown, S-Flow provides a declarative interface where users do not specify ``how'' to find the video snippets of interest;
instead, they only need to describe ``what'' their video snippets of interest look like, and Spatialyze will optimize accordingly.

From the users' perspective, Spatialyze automatically tracks objects from videos,
right after each of them is added to the {\small \tt world}.
Hence, they can filter the {\small \tt world} using these tracked objects.
Internally, Spatialyze does not execute any part of the workflow until
users call {\small \tt saveVideos()}.
Deferring executions this way allows Spatialyze to analyze the entire workflow for efficient execution.

We next discuss Spatialyze in detail (in \autoref{sec:execution}) and how it optimizes workflow execution (in \autoref{sec:optimization})
by leveraging videos' geospatial metadata and the physical behavior of real-world objects.
\aboveCodeVSpace%
\begin{lstlisting}[mathescape=true, escapeinside=!!, caption=\textmd{Geospatial {Video} {Analytic} {Workflow} {with} {Spatialyze}}, label=listing:python-workflow]
world = World()  # OR World(detect=CstmDetector, track=CstmTracker)  $\label{line:world}$
world.addGeogConstructs(RoadNetwork('road-network/boston-seaport/'))!$\label{line:road}$!
world.addVideo(GeospatialVideo(Video('v0.mp4'), Camera('c0.json')))  !$\label{line:gsvideo}$!
world.addVideo(GeospatialVideo(Video('v1.mp4'), Camera('c1.json')))  !$\label{line:end-build}$!
obj, cam, intersection = ( world.object(), world.camera(),!$\label{line:object}$!
                           world.geogConstruct(type='intersection')    )
world.filter( ((obj.type == 'car') | (obj.type == 'truck'))!$\label{line:filter}$!!$\label{line:pred-type}$!
            & (distance(obj, cam) < 50) & contains(intersection, obj)    !$\label{line:pred-distance}$!  !$\label{line:pred-contain}$!
            & headingDiff(obj, cam, between=[135, 225])                ) !$\label{line:pred-heading}$!  
world.saveVideos('output_videos/', addBoundingBoxes=True)!$\label{line:observe}$!
\end{lstlisting}
\belowCodeVSpace%

\section{Constructing Geospatial Video Analytic Workflows}
\label{sec:ql}

Spatialyze comes with S-Flow,
a DSL embedded in Python for users to construct their geospatial video data workflow.
In this section,
we describe S-Flow with Spatialyze's conceptual data model.

\subsection{Conceptual Data Model}
\label{sec:conceptual-data-model}
Analyzing videos in a general-purpose programming language requires users to manually extract and manipulate object bounding boxes from video frames and join these with geospatial metadata, which is tedious.
Hence, we developed Spatialyze's data model, a higher-level abstraction, to simplify user interaction with geospatial video.
This model includes 3 key concepts: 1) \emph{\cWorld},
2) \emph{\cGeoConstructs},
and
3) \emph{\cObjects},
each discussed further below.

\subsubsection{\cWorld}
\label{sec:cdl-world}
As discussed in \autoref{sec:example} and in line~\ref{line:world} of \autoref{listing:python-workflow},
users construct a \emph{\sWorld}
as a first step of
S-Flow.
Inspired by VisualWorldDB, a \emph{\cWorld} represents a
geospatial virtual environment which encompasses
\emph{\cGeoConstructs}
and
\emph{\cObjects},
all coexisting with spatial relationships.
These relationships, combined with the S-Flow,
enable users to describe
the videos of interest effectively. 

The \emph{\cWorld} also reflects the physical world, where \emph{\cObjects} exhibit realistic behaviors, like cars following lanes and adhering to speed limits.
We refer to these phenomena as {inherited physical behaviors} of \emph{\cObjects}.
We will discuss how Spatialyze leverages these
behaviors to optimize video processing in \autoref{sec:optimization}.

\subsubsection{\cGeoConstructs}
\label{sec:cdl-geog}
Each \emph{\cWorld} has a set of \emph{\cGeoConstructs};
each \emph{\cGeoConstruct} has a spatial property that defines its area, represented as a polygon.
It also has non-spatial properties including construct ID and construct type. 
We model each \emph{\cGeoConstruct} as
$(cid,\allowbreak type,\allowbreak ((x_0, y_0),\allowbreak (x_1, y_1),\allowbreak ...))$ where
$cid$ is its identifier;
$type$ is the construct type
(e.g., lane, intersection);
and $x_i, y_i$ is a vertex of the polygon that represents the construct.

\subsubsection{\cObjects}
\label{sec:cdl-object}
Besides static properties like \emph{\cGeoConstructs},
a \emph{\cWorld} may contain multiple \emph{\cObjects}.
Being movable,
each had spatiotemporal properties, e.g., its location at a given time.
Each also has non-spatiotemporal properties, including object ID and object type, that do not change throughout the object's life span.
For example, a \emph{\cObject} with object type of ``car'' will always be a ``car.''
Spatialyze models each \emph{\cObject} as
$(oid, type, ((l_0, r_0, p_0, t_0),\allowbreak(l_1, r_1, p_1, t_1), ...))$, where
$oid$ is its identifier;
$type$ is the object type, such as car, truck, or human.
At time $t_i$,
$l_i=(x_i, y_i, z_i)$ is its location;
$r_i$ is its 3D rotation, defined as a quaternion~\cite{kuipers:quaternion};
$p_i$ is its type-specific properties.

A {\em\cObject} with $type=camera$ represents a camera that captures videos.
It has a type-specific
property $p_i=(it_i,)$,
where $it_i$ is the intrinsic~\cite{anwar:camera,zhang:camera} of the camera at time $t_i$.
Despite being {\em\cObjects}, this paper distinguishes the camera from other types of {\em\cObjects} for the following reason.
{\em\cObjects} with $type=camera$ are parts of users' inputs into the {\em\cWorld}, while
Spatialyze infers {\em\cObjects} with other types from
the visual contents captured by the cameras.

We do not include ``video'' as a part of our conceptual data model.
A video itself does not have any geospatial property,
so it does not exist in a \emph{\cWorld}.
Similarly, none of the objects tracked in the video exists initially in the {\em\cWorld}.
Internally only after we bind the video with a camera
that captures the video,
the video gains geospatial properties from the camera and becomes a part of it.
Objects tracked in the video also gain their geospatial properties, as a result of the binding,
becoming a \emph{\cObject} in the \emph{\cWorld}.
Spatialyze derives the objects'
geospatial locations by combining their 2D bounding boxes and the camera's geospatial information.

\subsection{S-Flow}
\label{sec:our-syntax}

We design S-Flow based on the data model presented in \autoref{sec:conceptual-data-model}, where users interact and manipulate geospatial objects via \emph{\cGeoConstructs}
and \emph{\cObjects},
rather than video frames.
S-Flow's language constructs consist of functions that operate on \emph{\sWorld}, \emph{\sRoadNetwork}, and \emph{\sGSVideo}. %
Users compose such functions to construct geospatial video analytics workflows.

\subsubsection{\sCamera}
Lines~\ref{line:gsvideo}-\ref{line:end-build} of \autoref{listing:python-workflow} 
defines a {\em\cObject} with $type=camera$ as discussed in
in \autoref{sec:cdl-object}.
Users initialize a \emph{\sCamera} from a list, %
where the $i$-th element of the list is a dictionary with 4 fields: {\small \tt translation}, {\small \tt rotation}, {\small \tt intrinsic}, and {\small \tt timestamp}. 
They correspond to $l_i$, $r_i$, $it_i$, and $t_i$, as defined in \autoref{sec:cdl-object}.

\subsubsection{\sGSVideo}
A \emph{\sGSVideo} is a video enhanced with geospatial metadata.
A \emph{\sVideo} (lines~\ref{line:gsvideo}-\ref{line:end-build} of \autoref{listing:python-workflow}) by itself contains a sequence of images where each represents a frame of the video.
Such video exists spatially in the \emph{\cWorld} after 
its \emph{\sVideo} object is bound to a \emph{\sCamera} to obtain its spatial properties,
where each frame $i$ is an image taken from the camera at location $l_i$, rotation $r_i$,
intrinsic $it_i$, and at time $t_i$.
An object in each frame obtains its geospatial properties after its track is inferred 
from \emph{\sGSVideo}.
These objects then become \emph{\cObjects} as defined in \autoref{sec:cdl-object}.

\subsubsection{\sRoadNetwork}
\label{sec:road-network}
A \emph{\sRoadNetwork} is a collection of road segments.
Each road segment is a \emph{\cGeoConstruct} as defined in \autoref{sec:cdl-geog}.
Besides being a \emph{\cGeoConstruct}, a road segment stores its segment heading,
which indicates traffic headings in that road segment.
A road segment can have multiple segment headings (curve lane) or no segment heading (intersection).
Road segments in the \emph{\sRoadNetwork} dataset that we use have the following road types: 
roadsection,
intersection,
lane,
and lanegroup.
In S-Flow, we initialize a \emph{\sRoadNetwork} by loading from a directory containing files of the road network from each type.
This road network information is accessible and often provided with AV video datasets~\cite{fritsch:kitti-road, caesar:nuscenes}.

\subsubsection{\sWorld}
\label{sec:syntax-world}
Each language construct we introduced so far represents users' data.
Bringing the concept of \emph{\cWorld} described in \autoref{sec:cdl-world} into our DSL,
users interact with all of their video data through \emph{\cGeoConstruct} and \emph{\cObjects}
that exists in a \emph{\cWorld}.
In a Spatialyze workflow, users interact with the \emph{\sWorld} with this 3-step paradigm: {\em build}, {\em filter}, and {\em observe}.
First, users \textit{build} a \emph{\sWorld} by adding videos and geospatial metadata.
Once built, users \textit{filter} the \emph{\sWorld} to only contain their objects of interest.
Finally, users \textit{observe} the remaining objects through annotated videos.%

{\em Build.}
A \emph{\sWorld} is initially empty.
Users add \emph{\sRoadNetwork} using {\small \tt addGeogContructs()},
followed by adding \emph{\sGSVideo} via {\small \tt addVideo()}.
At this point of the workflow, the \emph{\sWorld} contains \emph{\cGeoConstructs},
and \emph{\sCameras}.
Spatialyze uses YOLOv5~\cite{josher:yolov5} and StrongSORT~\cite{du:strongsort} for object detection and
object tracking by default.
Spatialyze also supports using UDFs for object detection or tracking, shown as the comment in line~\ref{line:world} of \autoref{listing:python-workflow}.

{\em Filter.}
Users then construct predicates to filter objects of their interest.
As discussed in \autoref{sec:example}, Spatialyze has not extracted detections and trackings of objects from the videos yet at this point.
Nonetheless, Spatialyze provides an interface for users to interact with these objects as if they already exist.
For instance, {\small \tt object()} refers to an arbitrary \emph{\cObject}
with $type\ne camera$
in the \emph{\cWorld}
to use in the predicate.
Calling multiple {\small \tt object()} gives users multiple arbitrary {\em\cObjects}
that
appear
together in the same video.
For existing information,
{\small \tt camera()} refers to an arbitrary {\em\sCamera}.
Similarly, {\small \tt geoConstruct(type=...)} refers to a \emph{\cGeoConstruct} of a certain type.
Users can
construct a predicate involving multiple {\em\cObjects}, a {\em\sCamera},
or {\em\cGeoConstructs}, using our provided helper functions.
For instance, {\small \tt contains(intersection,obj)} determines if {\small \tt intersection} as a polygon contains {\small \tt obj} as a point.
{\small \tt headingDiff\allowbreak(obj,\allowbreak cam,\allowbreak between=\allowbreak[0,\allowbreak 9])} determines if the difference of the heading of {\small \tt obj} and the heading of {\small \tt cam} is between 0 and 9 degrees.
In addition, we provide more helper functions and an interface for users to define their own helper functions.
Finally, users can chain the {\small \tt filter} or use boolean operators to combine predicates.

{\em Observe.}
After filtering their desired \emph{\cObjects},
users observe the filtered \emph{\sWorld},
using one of the following 2 observer functions.
First, they can observe the \emph{\sWorld} by watching through snippets of the users' own input videos.
Using {\small \tt saveVideos\allowbreak(file,\allowbreak addBoundingBoxes)} function, Spatialyze saves all the video snippets that contain the \emph{\cObjects} of interest to a {\tt file},
where {\small \tt addBoundingBoxes} is a Boolean flag that indicates whether the bounding box of each \emph{\cObjects} should be shown.
Second, they can observe the \emph{\sWorld} by getting the \emph{\cObjects} directly.
Using {\small \tt getObjects()}, Spatialyze returns a list of \emph{\cObjects}.

\section{Workflow Execution in Spatialyze}
\label{sec:execution}

\begin{figure}%
  \centering%
  \includegraphics[width=\columnwidth]{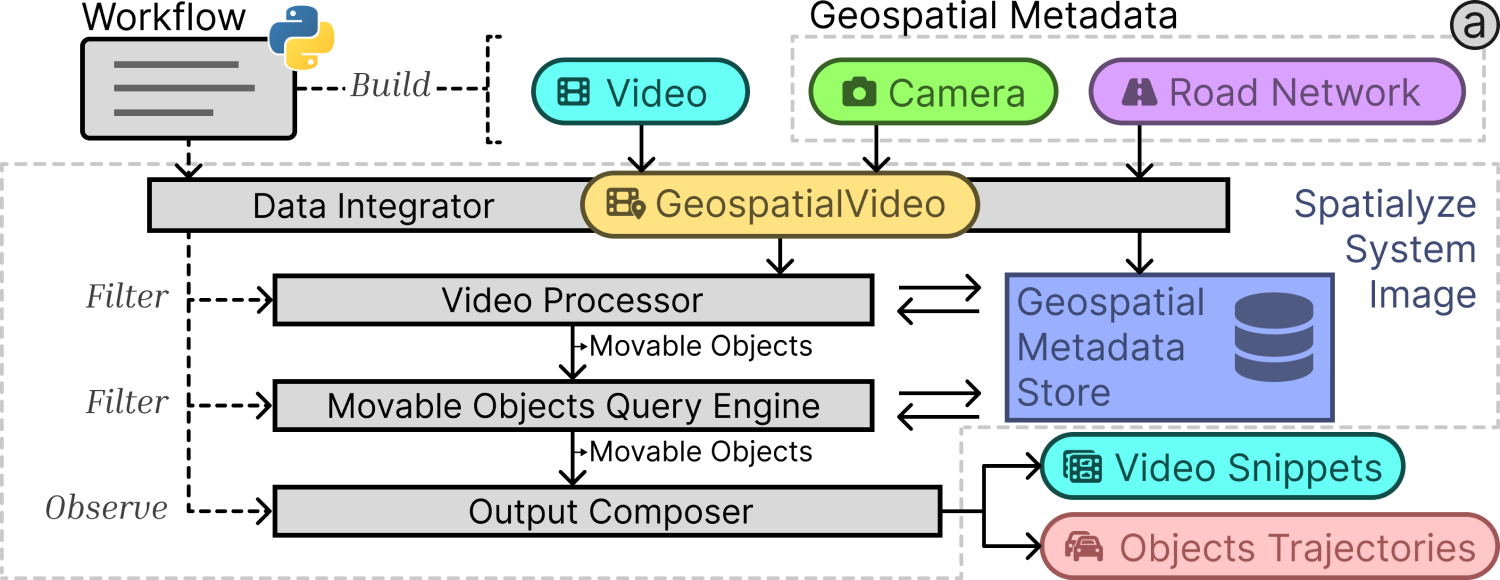}%
  \vspace{5pt}
  \includegraphics[width=\columnwidth]{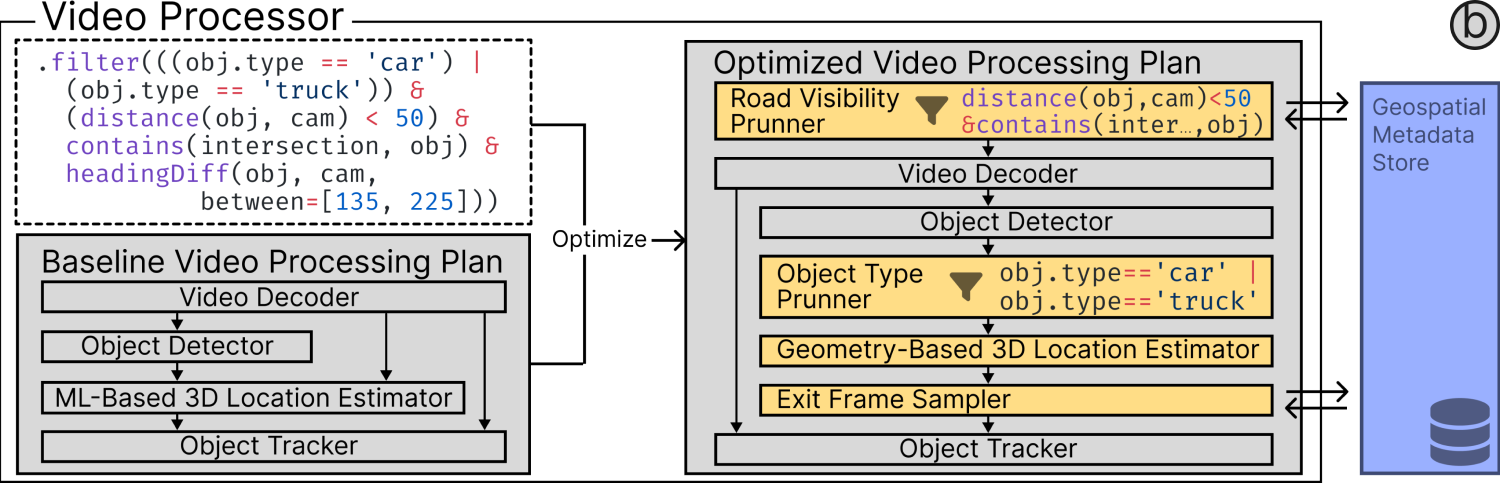}%
  \aboveCapVSpaceTwo%
  \Description{Spatialyze System Image: geospatial video analytic workflow execution.
  The user writes an S-Flow workflow in Python along with the input videos, cameras, and road network.
  Spatialyze's Data Integrator integrates the videos and cameras into geospatial videos and ingests the road network into the geospatial metadata store.
  Its Video Processor processes the geospatial videos to get the \cObjects.
  Its \cObjects Query Engine filters the relevant \cObjects based on the user's filter.
  Finally, its Output Composer composes either the videos with the relevant \cObjects or Object Trajectories for the user.
  
  Spatialyze optimizes a video processing plan by inserting optimization operators into the plan.
  The Road Visibility Pruner is before the Video Decoder.
  The Object Type Pruner is right after the Object Detector.
  The Geometry-Based 3D Location Estimator replaces the ML-based 3D Location Estimator.
  The Exit Frame Sampler is right before the Object Tracker.}%
  \caption{\bigcircled{a} Geospatial Video Analytics Workflow Execution.
  \bigcircled{b} Processing Plan for the workflow shown in \autoref{listing:python-workflow}.}\belowCapVSpace%
  \label{fig:system-arch}%
  \label{fig:video-processing}%
\end{figure}%

From users' view, Spatialyze
filters objects in the \emph{\sWorld} after they call {\small \tt filter()}.
Then, they can \emph{observe} the filtered objects afterward,
using one of the observer functions described in \autoref{sec:syntax-world}.
Internally, Spatialyze only performs video processing to detect objects and compute tracks for them {\em as needed} given the workflow.
For example, in line~\ref{line:pred-heading} of \autoref{listing:python-workflow}, the user filters on objects' moving directions.
Executing this requires computing the trajectory of objects in the videos to compute their directions.
Given an S-Flow workflow,
the goal of Spatialyze
is to understand the users' workflow,
recognize and execute only the video processing operators needed for users' filter predicates,
and return \emph{\cObjects} that satisfy the filters.

In \autoref{sec:ql}, we discuss a high-level interface for Spatialyze that allows users to construct their geospatial video analytics workflows declaratively.
Users follow the \emph{build}-\emph{filter}-\emph{observe} pattern without knowing when the actual execution happens.
Internally, all execution happens {\em when the users observe the world}.
In this section, we discuss the challenges in designing Spatialyze system
and how Spatialyze generates execution plans for a workflow.

\subsection{Challenges in Workflow Execution}
\label{sec:sys-challenge}
We identify 3 challenges in building a geospatial video analytics system,
specifically designing a system that is fast, capable of processing arbitrary-length videos, and extensible.
First, processing videos with ML functions is slow;
Spatialyze must leverage its knowledge of the entire workflow to decide which video frames and ML functions must be executed.
Second, one minute of a 1080p 12fps decoded video is 4.4 GB.
Two days of such video can already fill up one GCP VM with the largest available memory size of 11,776 GB as of 2023.
Spatialyze must scale with video sizes with limited memory.
Finally,
Spatialyze must be extensible with future ML functions and optimization techniques.
We discuss how we address these challenges in \autoref{sec:execution-stages} and \autoref{sec:optimization}.

\subsection{System Design \& Workflow Execution}
\label{sec:execution-stages}

To minimize execution time,
Spatialyze defers
ML inferences, which are expensive,
until users \emph{observe} the \emph{\cWorld}.
This way, Spatialyze can optimize the entire workflow execution.
Once users {\em observe} a {\em\sWorld},
Spatialyze starts executing users' workflow in 4 stages sequentially:
\circled{1} Data Integrator,
\circled{2} Video Processor,
\circled{3} \cObjects Query Engine,
and \circled{4} Output Composer;
as shown in \autoref{fig:system-arch}.

\subsubsection{Data Integrator}
\label{sec:data-integrator}
Spatialyze integrates the input data
when users \emph{Build} their \emph{\cWorld}.
It processes the \emph{\cGeoConstructs} by creating tables in a geospatial metadata store
to store the \emph{\sRoadNetwork} and create spatial indexing for every \emph{\cGeoConstruct}.
Then, it joins each \emph{\sVideo} and its corresponding \emph{\sCamera} frame-by-frame,
by their frame number,
resulting in a {\em\sGSVideo}.

\subsubsection{Video Processor}
\label{sec:video-processing}
This stage takes in users' \emph{\sGSVideo} and users' filter predicates from their workflow's \emph{Filter}.
It then extracts \emph{\cObjects} from the \emph{\sGSVideo}.
To do so, it first decodes the video to get image frames.
It then executes an object detector to extract bounding boxes of objects from each frame.
Based on the bounding box, image frame, and {\em\sCamera}, it estimates the 3D location of each object.
Finally, it executes the object tracker on the bounding boxes and image frame to get {\em\cObjects}.
By default, Spatialyze uses an off-the-shelf algorithm to implement each step.
The video decoder is implemented using OpenCV~\cite{bradski:opencv} as it is the fastest option available.
The object detector is implemented using YOLO~\cite{josher:yolov5},
a SOTA object detector used by our prior work~\cite{romero:viva,xu:eva} with high accuracy, low runtime, and capability to detect multiple types of objects.
The object 3D location estimator is implemented using Monodepth2~\cite{godard:monodepth} as it is the SOTA image depth estimation that only requires an image as an input.
The object tracker is implemented using StrongSORT~\cite{du:strongsort} as it is a fast and online method,
suited for processing large videos.

We designed the video processor to stream video frames through all the above algorithms.
Each video frame is pipelined through decoding, object detection, 3D location estimation, and object tracking; hence our video processor only needs to keep $O(1)$ video frames
and can handle videos of arbitrary size, addressing the second challenge.
While stream processing of videos has been used previously~\cite{brostrom:yolo-strongsort},
applying it to a streaming video processor is not trivial.
For example, the original YOLO-StrongSORT object tracker~\cite{brostrom:yolo-strongsort} streams video frames.
However, naively extending the tracker with our optimization techniques requires
hard-coding them into the tracker's implementation.
This does not scale to future trackers.
To solve this problem (and address the third challenge),
we implemented our video processor as a plan of streaming operators shown in \autoref{listing:baseline-processing-plan}.
Our baseline operators\footnote{
The {\bf streaming operators} internally process videos and geospatial metadata to extract (2D or 3D) object detections and/or tracks.
In contrast, users use {\bf predicate operators} (e.g., {\scriptsize\tt contains}, {\scriptsize\tt distance}) to define filter predicates.
}
include:
\circled{1} Video Decoder;
\circled{2} Object Detector;
\circled{3} 3D Location Estimator;
and \circled{4} Object Tracker.
Specifically, \circled{1} decodes each video to get a sequence of RGB images representing video frames.
\circled{2} detects objects in each frame from \circled{1},
returning each object and its object type and 2D bounding box.
A\emph{\sWorld} is 3 dimensional where all objects have 3D locations;
therefore, \circled{3} estimates the 3D location of each object from \circled{2}
based on its \emph{\sCamera}.
Lastly, \circled{4} tracks objects
by associating detections from \circled{3} between frames.
We implemented each operator as an iterator function,
where each operator takes in streams of per-frame inputs and returns a stream of per-frame outputs.
As a result, our video processor can process videos of arbitrary size,
including those larger than the available memory.
In addition,
we implemented a caching system for the operators.
In \autoref{listing:baseline-processing-plan},
{\small\tt frames} is an input to both {\small\tt ObjectDetector} and {\small\tt ObjectTracker}.
Instead of computing the stream of {\small\tt frames} twice,
Spatialyze iterates through {\small\tt frames} once and caches the results.
For instance, a frame $i$ is decoded when the {\small\tt ObjectDetector} processes it.
Spatialyze caches the frame $i$ so that the {\small\tt ObjectTracker} can process it without having to decode the frame again.
Spatialyze keeps track of the number of the operators that will be using
a cached result and evicts a frame as soon as all of the operators finished processing it;
therefore, Spatialyze only caches a few frames at a given time.

Stateless operators like {\small\tt ObjectDetector} or {\small\tt Loc3DEstm} process each video frame independently.
In contrast, {\small\tt ObjectTracker} is a stateful operator that keeps track of what objects have been tracked so far at a given frame.
To retain states, an ML function can be implemented as a stateful streaming operator using our provided wrapper with minimal changes to the original code of the function.

\vspace{-3pt}
\aboveCodeVSpace%
\begin{lstlisting}[language=Python, mathescape=true, escapeinside=!!, caption=\textmd{Baseline Video Processing Plan}, label=listing:baseline-processing-plan]
frames          = VideoDecoder('videofile.mp4')           !$\label{line:plan-decode}$!
object2Ds       = ObjectDetector(frames)                  !$\label{line:plan-detect}$!
object2DAnd3Ds  = Loc3DEstm(frames, object2Ds, cameras)   !$\label{line:plan-3d}$!
movableObjects  = ObjectTracker(object2DAnd3Ds, frames)   !$\label{line:plan-track}$!
\end{lstlisting}
\belowCodeVSpace%
\vspace{-3pt}

Executing the plan computes 3D tracks of objects in a video.
However, streaming each video through all operators is computationally expensive and not always necessary.
Based on the users' filter predicates from when they \emph{Filter} the \emph{\cWorld},
Spatialyze only includes the necessary operators in the execution plan.
For instance, a predicate based on object types like {\small \tt obj.type=="car"} requires
objects' types and hence the execution plan will include lines \ref{line:plan-decode} and \ref{line:plan-detect}.
Likewise, a predicate with {\small \tt distance} or {\small \tt contain} functions requires
objects' 3D location and therefore
include lines \ref{line:plan-decode}, \ref{line:plan-detect}, and \ref{line:plan-3d} in the plan.
Finally, a predicate with a {\small \tt headingDiff} function requires
objects' moving directions and include 
all lines in the plan.

We implemented each optimization technique in \autoref{sec:optimization} as a streaming operator.
Two of the optimization operators execute users' predicate early on to
reduce the number of inputs (video frames and objects) into later operators.
We designed our video processor that constructs plans using streaming operators;
therefore,
future optimization techniques and video processing functions can be added,
addressing the runtime efficiency and extensibility challenges.

\subsubsection{\cObjects Query Engine}
\label{sec:query-engine}
This stage takes in users' filter predicates from their workflow's \emph{Filter} and resulting \emph{\cObjects} from the video processor.
It then filters the \emph{\cObjects} with the users' predicates in 2 steps.
First, it streams the \emph{\cObjects} into a newly created table in the geospatial metadata store.
The table has a spatial index on objects' tracks and a temporal index on the period, in which the objects exist.
Then, it translates the user's filter predicates into an SQL query
and executes it in the geospatial metadata store against the \emph{\cObjects}, {\em\sCameras}, and \emph{\sRoadNetwork}.
When the user's predicates involve multiple objects (e.g., {\footnotesize\tt distance(car1,car2)<5}),
Spatialyze self-joins the {\em\cObjects} table to find all objects that coexist in a given period using the temporal index to speed up the join.
When the predicates involve spatial relationships between a {\em\cObject} and a {\em\sRoadNetwork} (e.g., {\footnotesize\tt contains(road,car)}),
the spatial index
speeds
up the joins between the {\em\cObjects} and {\em\sRoadNetwork} tables.
We implemented the metadata store using MobilityDB~\cite{zimanyi:mobility}
as it provides necessary spatial and temporal indexes for movement objects.
The index along with its spatiotemporal data type are suited for
processing queries on {\em\cObjects} and {\em\cGeoConstructs}.

\subsubsection{Output Composer}
\label{sec:output-composer}
This stage composes the \emph{\cObjects} query results to users' preferred formats for them to \emph{observe}.
Calling {\small \tt getObjects}, Spatialyze returns a list of \emph{\cObjects}.
Calling {\small \tt saveVideos}, 
Spatialyze encodes the frames containing the \emph{\cObjects} into video files with colored bounding boxes annotation. %

\autoref{fig:system-arch}
shows the execution stages for the workflow in~\autoref{listing:python-workflow}.
It contains all the stages mentioned above.
A user first \emph{builds} the world by adding data.
Spatialyze's Data Integrator creates an indexed table consisting of \emph{\sRoadNetwork} and
joins the \emph{\sVideo} and its corresponding \emph{\sCamera}.
Then, Spatialyze's Video Processor creates an optimized video processing plan with 4 processing operators and 4 optimization operators (to be discussed in \autoref{sec:optimization}).
Parts of the users' predicates are executed in \emph{\inview} and \emph{\objtype}.
The \cObjects Query Engine streams the \emph{\cObjects} from the Video Processor into the geospatial metadata store
and filters them with the users' predicates.
Finally, as the workflow asks for videos, Spatialyze formats and saves the frames that have the filtered objects into video snippet files.%

\section{Video Processing Optimization}
\label{sec:optimization}
SOTA
video inference uses ML models to extract the desired information,
such as objects' positions and trajectories within each video.
Running such models is computationally expensive.
Spatialyze's video processor operates with 4 optimization techniques to reduce such runtime bottlenecks.
To be discussed below, the idea behind our optimization is to leverage the geospatial properties of the videos,
and the \emph{\cObjects}' {inherited physical behaviors}.
We implement each optimization as a streaming operator in the Video Processing stage in \autoref{sec:video-processing}.
In addition to the existing 4 processing operators,
Spatialyze may place optimization operators in the middle of the video processing plan to reduce the amount of input into later processing operators based on the filter predicates.
{\bf In \autoref{sec:inview}}, Spatialyze places the \emph{\inview} after the Video Decoder,
when users specify the \emph{\cGeoConstructs} the objects of interest must be in.
Doing so reduces 
the number of input frames for the rest of the operators in the plan.
{\bf In \autoref{sec:objtype}}, If users filter on object types, then Spatialyze places the \emph{\objtype} right after the Object Detector.
It reduces the number of objects that Spatialyze needs to calculate for their 3D locations and tracking.
{\bf In \autoref{sec:depth}}, Spatialyze replaces the ML-based 3D Location Estimation operator with the much faster \emph{\depthEstimation},
if Spatialyze can assume that the objects of interest are on the ground, such as cars or people.
{\bf In \autoref{sec:sample}}, Spatialyze places the \emph{\sample} in between the 3D Location Estimator and the Object Tracker,
if the users' predicates focus on cars.
It utilizes the 3D locations of the objects and the user-provided metadata to reduce the number of frames that the Object Tracker processes. %

\autoref{fig:video-processing} shows where the video processor executes all the optimization steps.
Line~\ref{line:pred-contain} in \autoref{listing:python-workflow} corresponds to \emph{\inview}.
Line~\ref{line:pred-type} filters by object type, so Spatialyze adds \emph{\objtype}.
The objects of interest are cars or trucks, so Spatialyze adds \emph{\sample} and \emph{\depthEstimation}.
In the following sections, we discuss each optimization in detail.

\subsection{\inview}
\label{sec:inview}
In our experiments, the runtime of ML models~\cite{josher:yolov5, zhou:osnet, du:strongsort} in the video processing plan
takes \bmv{\evalVariableZeroVProcessPercent\%} of the entire workflow execution runtime, on average.
Hence, 
the goal of the \inview is to remove frames that do not contain objects of the user's interest to reduce the overall runtime of the video processing plan.

\subsubsection{High-Level Concepts}
The \emph{\inview} uses \emph{\cGeoConstruct}'s visibility as a proxy for a \emph{\cObject}'s visibility.
Specifically, a predicate includes {\small \tt contains\allowbreak(road\allowbreak,obj)} and {\small \tt distance\allowbreak(cam\allowbreak,obj)<d} means that {\small \tt obj} is not visible in the frames where {\small \tt road} is not visible within {\small \tt d} meters;
therefore, Spatialyze prunes out those frames.
In \autoref{listing:python-workflow},
the \emph{\inview} partially executes lines~\ref{line:pred-contain}
where the user searches for ``cars within 50 meters at an intersection.''
Spatialyze removes all frames that do not contain any visible intersection within 50 meters
as these frames will not contain any car of users' interest.

The \emph{\inview} only concerns the visibility of \emph{\cGeoConstructs}.
Therefore, it uses \emph{\cGeoConstructs} and \emph{\sCamera} to decide whether to prune out a video frame.
It uses the intrinsic and extrinsic properties of \emph{\sCamera}~\cite{zhang:camera} to determine the viewable area of each of the cameras according to the world coordinate system.\footnote{\emph{World Coordinate System} is a 3-dimensional Euclidean space, representing a \emph{\cWorld}.}
It then determines if the viewable area of each frame includes any \emph{\cGeoConstruct} of interest; 
frames that do not contain such constructs are dropped and will not be processed in the later processing operators.
Since the \emph{\inview} will be the first processing operator if applied,
it can reduce the number of frames that need to be processed by subsequent ML models in the plan,
and hence can substantially reduce overall execution time.

\subsubsection{Algorithm}
\label{sec:inview-algo}
The {\em \inview} works on a per-frame basis.
For each video frame $i$ shot by a \emph{\sCamera} at location $l_i$ with 
rotation $r_i$ and camera intrinsic $it_i$,
the \emph{\inview} performs three steps to determine if the frame should be pruned out.
First, it computes the 3D viewable space of the \emph{\sCamera} at frame $i$.
Then, it uses the viewable space to find \emph{\cGeoConstructs} that are visible to the \emph{\sCamera} at frame $i$.
Finally, it uses the users' filter predicates to determine if any \emph{\cGeoConstruct} of interest is visible in the frame.
We explain each step below.

\begin{figure}%
  \centering%
  \includegraphics[width=\columnwidth]{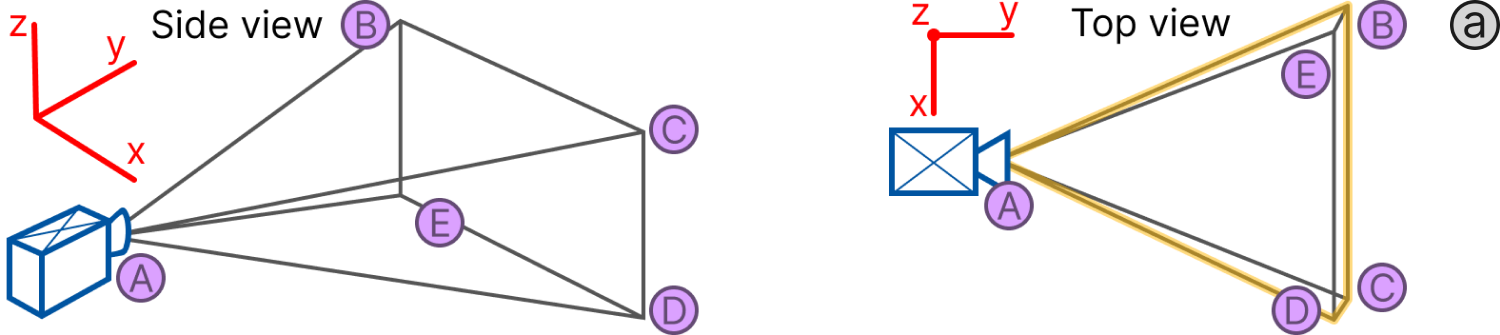}%
  \vspace{3pt}
  \includegraphics[width=\columnwidth]{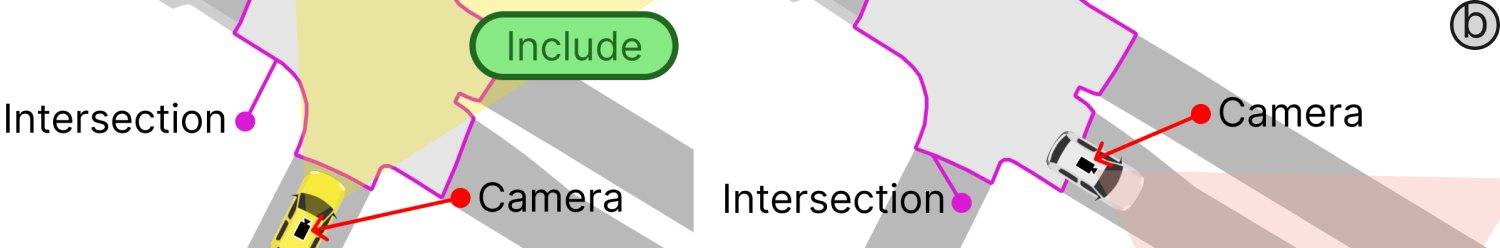}%
  \Description{The illustrations of the side and top view of the camera's viewable space (in pyramid shape), where points A represents the camera position, and points B, C, D, and E represent the top-left, top-right, bottom-right, bottom-left of the frame at d meters in front of the camera.
  The left figure shows the camera's viewable area intersecting with a polygon representing a road intersection.
  The right figure shows the camera's viewable area, which does not intersect with any polygon representing a road intersection.}%
  \aboveCapVSpaceTwo\caption{%
  \bigcircled{a} A 3D viewable space of a camera in a pyramid shape and
  its projected area of the pyramid onto the $z=0$ plane.
  \bigcircled{b} In \autoref{listing:python-workflow},
  the video processor executes expensive ML models only on video frames with a visible intersection.%
  }\belowCapVSpace%
  \label{fig:camera-view}%
\end{figure}

First, the \emph{\inview} calculates the 3D viewable space of the \emph{\sCamera} at frame $i$.
The viewable space is a pyramid (\autoref{fig:camera-view} side view).
\circled{{A}} is the location of the camera.
\circled{{B}}, \circled{{C}}, \circled{{D}}, and \circled{{E}} represent the top-left, top-right, bottom-right, and bottom-left of the frame at $d$ meters in front of the camera.
We compute \circled{{B}}, \circled{{C}}, \circled{{D}}, and \circled{{E}} by converting the mentioned 4 corner points of the frame from pixel (2D) to world (3D) coordinates.
This is done in 3 steps by converting a point from:
1) \emph{Pixel to Camera Coordinate System};
2) \emph{Camera to World Coordinate System};
and 3) \emph{Pixel to World Coordinate System}.
\eqspace

{\em Pixel to Camera Coordinate System.}
The equation for converting a 3D point $(x_c, y_c, z_c)$ from camera coordinate system\footnote{
    \emph{Camera Coordinate System} is a 3-dimensional Euclidean space from the perspective of a camera.
    The origin point is at the camera position.
    Z-axis and X-axis point toward the front and right of the camera.
    The Y-axis points downward from the camera.
} to a 2D point $(x_p, y_p)$ in pixel coordinate system is:
\begin{equation} \label{eq:c2p}
    \left[\begin{smallmatrix}
        x_p \\ y_p \\ 1
    \end{smallmatrix}\right] = 
    \left[\begin{smallmatrix}
        f_x & s & x_0 \\ 0 & f_y & y_0 \\ 0 & 0 & 1
    \end{smallmatrix}\right] \times
    \left[\begin{smallmatrix}
        x_c \\ y_c \\ z_c
    \end{smallmatrix}\right] \cdot \frac{1}{z_c},
    \text{where } it_i = \left[\begin{smallmatrix} f_x & s & x_0 \\ 0 & f_y & y_0 \\ 0 & 0 & 1 \end{smallmatrix}\right]
\end{equation}
where $it_i$ represents the camera intrinsic,
a set of parameters representing {\em\sCamera}'s focal lengths ($f_x,f_y$), skew coefficient ($s$), and optical center in pixels ($x_0, y_0$).
The intrinsic is
specific to each camera,
independent of its location and rotation.
It converts 3D points from a camera perspective to 2D points in pixels.

To convert a 2D point in pixel ($x_p$ and $y_p$),
we can find the 3D point in the camera coordinate system ($x_c$, $y_c$, and $z_c$) if we know the distance of the point from the camera ($z_c$).
From \autoref{eq:c2p}, we have:
\begin{equation} \label{eq:_p2c}
\begin{split}
    \left[\begin{smallmatrix}
        x_c & y_c & z_c
    \end{smallmatrix}\right]^\top
    & = it_i^{-1} \times
    \left[\begin{smallmatrix}
        x_p & y_p & 1
    \end{smallmatrix}\right]^\top \cdot z_c %
\end{split}
\end{equation}
Modifying \autoref{eq:_p2c} so that the left side is $[x_c, y_c, z_c, 1]^\top$,
\begin{equation} \label{eq:p2c}
\begin{split}
    \left[\begin{smallmatrix}
        x_c \\ y_c \\ z_c \\ 1
    \end{smallmatrix}\right]
    & = \left[\begin{smallmatrix}
        \begin{array}{c|c}
            \text{\small\emph{\text{$it_i^{-1}$}}} & \text{\small{0}} \\
            \hline
            \text{\small{0}}         & \text{\small{1}}
        \end{array}
    \end{smallmatrix}\right] \times
    \left[\begin{smallmatrix}
        x_pz_c \\ y_pz_c \\ z_c \\ 1
    \end{smallmatrix}\right] \\[-8pt]
    & = C \times
    \left[\begin{smallmatrix}
        x_pz_c & y_pz_c & z_c & 1
    \end{smallmatrix}\right]^\top,
    \text{where } C = \left[\begin{smallmatrix}
        \begin{array}{c|c}
            \text{\small\emph{\text{$it_i^{-1}$}}} & \text{\small{0}} \\
            \hline
            \text{\small{0}}         & \text{\small{1}}
        \end{array}
    \end{smallmatrix}\right]
\end{split}
\end{equation}

{\em Camera to World Coordinate System.}
For this conversion, we set up $R$, a $3 \times 3$ matrix,
representing the 3D rotation of the camera that is derived from the camera's rotation quaternion.
And, $t$ is a 3D vector, representing the camera's translation.
Then, a $3 \times 4$ extrinsic matrix $[R | t]$ converts 3D points from camera coordinate system $(x_c, y_c, z_c)$ to 3D points $(x, y, z)$ in world coordinate system:
\begin{equation} \label{eq:c2w}
    \left[\begin{smallmatrix}
        x & y & z
    \end{smallmatrix}\right]^\top = [R | t] \times
    \left[\begin{smallmatrix}
        x_c & y_c & z_c & 1
    \end{smallmatrix}\right]^\top
\end{equation}

{\em Pixel to World Coordinate System.}
Combining the previous 2 conversions in \autoref{eq:p2c} and \autoref{eq:c2w}
converts 2D points from a pixel coordinate system to 3D points in the world coordinate system:
\begin{equation} \label{eq:p2w}
\begin{split}
    \left[\begin{smallmatrix}
        x & y & z
    \end{smallmatrix}\right]^\top
    & = [R | t] \times
    C \times
    \left[\begin{smallmatrix}
        x_pz_c & y_pz_c & z_c & 1
    \end{smallmatrix}\right]^\top
\end{split}
\end{equation}

{\em 4 corners of a video frame in the World Coordinate System.}
To convert the 4 corners of a video frame from pixel to world coordinates, we modify \autoref{eq:p2w} where
we substitute $z_c$ with $d$, and $x_p$ and $y_p$ with actual pixel values representing the 4 corners of the video frame;
$w$ and $h$ are the width and height in pixels of the video frame, and 
each of $B, C, D, E$ is a $3 \times 1$ vector representing its 3D location.
\begin{equation} \label{eq:ps2w}
    \left[\begin{smallmatrix}
        B & C & D & E
    \end{smallmatrix}\right] = [R | t] \times
    C \times
    \left[\begin{smallmatrix}
        0 & wd & wd & 0 \\
        0 & 0 & hd & hd \\
        d & d & d & d \\
        1 & 1 & 1 & 1
    \end{smallmatrix}\right]
\end{equation}

After the corners of each frame are translated to world coordinates, the \emph{\inview} uses the viewable space to determine the \emph{\cGeoConstructs} that are visible by \emph{\sCamera} at frame $i$.
Each \emph{\cGeoConstruct}'s area is defined using a 2D polygon.
Specifically, all polygons representing \emph{\cGeoConstructs} are 2D in the $z=0$ plane.
Therefore, the \emph{\inview} projects the computed 3D viewable space onto the $z=0$ plane.
We define the top-down 2D viewable area as the convex hull~\cite{sklansky:convexhull} of the projected vertices \circled{{A}}, \circled{{B}}, \circled{{C}}, \circled{{D}}, and \circled{{E}} (yellow highlight in \autoref{fig:camera-view}'s top view).
We then spatially join the 2D viewable area with {\em \cGeoConstructs} in the geospatial metadata store,
using their geospatial indexing created in \autoref{sec:data-integrator}.
Any \emph{\cGeoConstruct} that overlaps with the 2D viewable area is considered visible to the \emph{\sCamera} at frame $i$.
Once computed, we get {\small \tt visibleGeogs}, a set of \emph{\cGeoConstruct} types that are visible to the \emph{\sCamera} at frame $i$.

Finally, the \emph{\inview} uses the users' filter predicates to determine if there is any \emph{\cGeoConstruct} of interest visible in the frame.
Examining the {\small \tt contains} predicates in the filter,
it assigns each {\small \tt contains} to $True$ if its first argument is in {\small \tt visibleGeogs} and $False$ otherwise.
If the value of the transformed filter predicates is $False$, then the \emph{\inview} prunes out that frame.

\subsubsection{Benefits and Limitations}
The \emph{\inview} has an insignificant overhead runtime compared to the runtime it can save.
Our experiments show that the \emph{\inview} takes \bmv{\evalVariableOneInviewPercent\%} of the video processing runtime to execute while saving up to \bmv{\evalVariableOneInviewQueryTwoRuntimeReduction\%} of the video processing runtime.
So in the worst case that \emph{\inview} cannot prune out any video frames, the runtime increase is still negligible.
However, having the \emph{\inview} in the video processing plan may cause tracking accuracy to drop in some scenarios.
Using \autoref{sec:example} as an example, if a car starts at an intersection,
exits the intersection,
and enters another intersection.
When the car is not in any intersection,
the camera may not capture any frames containing an intersection.
The \emph{\inview} will prune out the period of video frames where no intersection is visible to the camera.
The object tracker might consider the car before leaving the intersection and the car after entering another intersection to be two different cars.

\subsection{\objtype}
\label{sec:objtype}

SORT-Family algorithms~\cite{du:strongsort, wojke:deepsort, bewley:sort} use the
Hungarian method~\cite{kuhn:hungarian} to associate objects between each 2 consecutive video frames.
The runtime of the Hungarian method scales with the number of objects to associate in each frame.
Reducing the number of objects input into the algorithms reduces the runtime of
the object tracker.

\subsubsection{High-Level Concepts}
Spatialyze only needs to track objects
of interest as stated in the workflow.
The \emph{\objtype} prunes out irrelevant objects from being tracked.
We analyze users' filter predicates to look for types of objects that are necessary for computing the workflow outputs.
Using the predicates in \autoref{listing:python-workflow} as an example,
the \emph{\objtype} partially executes the line~\ref{line:pred-type},
where users only look for cars and trucks in the workflow outputs.
The trajectories of other objects (\emph{e.g.}, humans%
) will not appear in the final workflow outputs anyway,
so tracking them is unnecessary.
Therefore, we only need to input objects
detected as cars or trucks into the object tracker to reduce its workload.

\subsubsection{Benefits and Limitations}
The \emph{\objtype}
reduces the object tracker's runtime.
It
has low overhead runtime of \bmv{\evalVariableTwoObjTypeQueryTwoRuntimePercent\%} and does not require geospatial metadata
but can reduce up to \bmv{\evalVariableZeroObjTypeQueryOneSSRuntimeReduction\%} of 
the tracking runtime,
which is \bmv{\evalVariableZeroObjTypeQueryOneRuntimeReduction\%} of the video processing time.

\subsection{\depthEstimation}
\label{sec:depth}

In our video processing plan, Spatialyze estimates an object's 3D location using its distance from the camera and calculated bounding box.
Specifically, we use YOLOv5~\cite{josher:yolov5} to estimate the bounding box and Monodepth2~\cite{godard:monodepth} to estimate the object's distance from the camera in our baseline plan.
Monodepth2 only requires monocular images to estimate each object's distance from the camera;
therefore, it is flexible in general use cases.
Spatialyze can always use Monodepth2 to estimate any object's distance from the camera in a video frame.
However, this algorithm is slow and can take up to \bmv{\evalVariableZeroMonodepthAllPercent\%} portion of the total baseline video processing runtime.
Therefore, we explored an alternative algorithm that leverages existing geospatial metadata to speed up 3D object location estimation.

\subsubsection{High-Level Concepts}
\emph{\depthEstimation} leverages \emph{\sCamera} to estimate an object's 3D locations from its 2D bounding box.
It is a substitution for Monodepth2 to estimate objects' 3D locations.
\emph{\depthEstimation} uses basic geometry calculations and requires 2 assumptions to hold.
First, the object of interest must be on the ground.
We can then assume that the lower horizontal border of the object's bounding box is where the object touches the ground.
Second, we can identify the equation of the plane that the object of interest is on.
In Spatialyze, the plane $z=0$ is the ground,
although our algorithm can be modified to be used with any plane.
The middle point of the lower border of each object's 2D bounding box represents the point that it touches the ground (red \cstar{red} in \autoref{fig:estimate-3d}).

\subsubsection{Algorithm}
\emph{\depthEstimation} performs 2 steps to estimate the 3D location of an object from its 2D bounding box.
First, based on the second assumption, \emph{\depthEstimation} finds an equation that represents a \emph{Vector of Possible 3D Locations} of an object.
Second, based on the first and second assumptions, \emph{\depthEstimation} finds an \emph{Exact 3D Location} from the vector that intersects with the ground.

\begin{figure}%
  \centering%
  \includegraphics[width=\columnwidth]{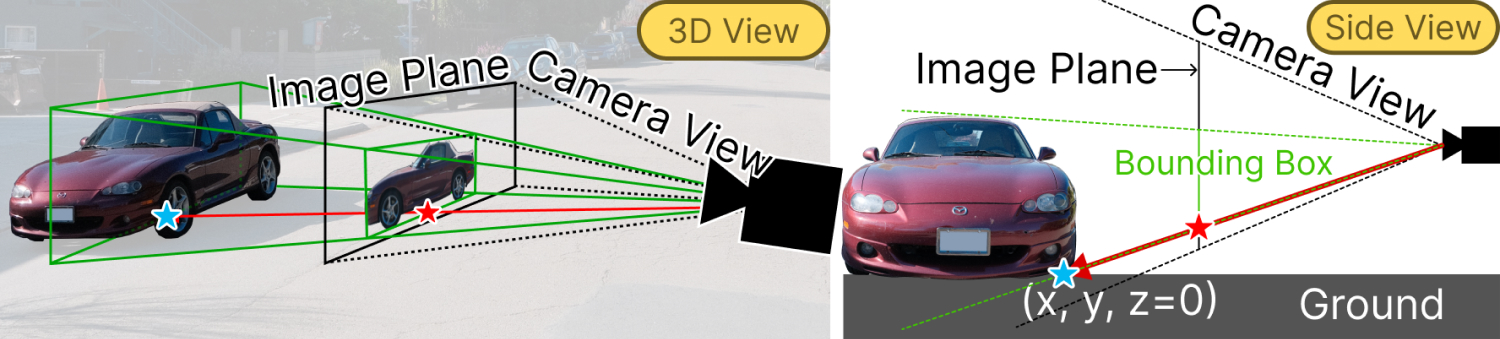}%
  \Description{A 3D illustration of geometry-based 3D location estimation, showing relationships between the camera, the 2D car in the image frame, and the actual car location}%
  \aboveCapVSpaceTwo\caption{From a car's 2D location (\cstar{red}) to its 3D location (\cstar{cyan}).}\belowCapVSpace%
  \label{fig:estimate-3d}%
\end{figure}

{\em Vector of Possible 3D Locations.}
Using \autoref{eq:p2w}, we convert any 2D point in the pixel coordinate system to a 3D point in the world coordinate system, if we know $z_c$.
Since we do not know $z_c$ yet,
we can derive a parametric equation representing all possible 3D points (red line in \autoref{fig:estimate-3d}) converted from the 2D point.
Let $d$ be the unknown distance of the object from the camera.
\begin{equation} \label{eq:p_p2w}
\begin{split}
    \left[\begin{smallmatrix}
        x & y & z
    \end{smallmatrix}\right]^\top
    & = [R | t] \times
    C \times
    \left[\begin{smallmatrix}
        x_pd & y_pd & d & 1
    \end{smallmatrix}\right]^\top
\end{split}
\end{equation}

{\em Exact 3D Location.}
We assume that the bottom line of the object's bounding box touches the ground;
therefore, $(x, y, z)$ also touches the ground.
Because the ground is $z=0$ in our case,
we can solve the \autoref{eq:p_p2w} for $d$.
Once we know $d$, we can solve the parametric \autoref{eq:p_p2w} for $x$ and $y$ (blue \cstar{cyan} in \autoref{fig:estimate-3d}).

\subsubsection{Benefits and Limitations}
\emph{\depthEstimation} relies on the assumptions mentioned, which do not always hold, especially, the first one.
Nevertheless, Spatialyze can analyze users' filter predicates and tell whether the types of objects of interest can be assumed to touch the ground, applying the optimization only then.
For example, ``car,'' ``bicycle,'' or ``human'' touch the ground, while ``traffic light'' does not.
In addition, \emph{\depthEstimation} is \bmv{\bmv{\evalVariableZeroDepthAllTDEstmSpeedup}$\times$} faster compared than Monodepth2 on average.
If one of the estimated 3D locations of an object in a video frame ends up being behind the camera,
which means that the object does not touch the ground,
Spatialyze can fall back to using Monodepth2 for that frame.
Otherwise, \emph{\depthEstimation} provides \bmv{\evalVariableTwoDepthAllTDEstmReductionPercent\%} runtime reduction on average for object 3D location estimation and \bmv{\evalVariableZeroDepthAllRuntimeReduction\%} for the video processing.

\subsection{\sample}
\label{sec:sample}

Common object tracking algorithms~\cite{du:strongsort,wojke:deepsort,bewley:sort},
including that used in Spatialyze in the object tracker (\autoref{sec:video-processing}),
performs tracking-by-detection~\cite{leal-taixe:track-by-detect}.
An object detector identifies objects in video frames but does not link detections across frames to represent the same object.
The object tracker associates detections in consecutive frames,
creating a chain of associated detections that forms an object trajectory, such as a {\em\cObject}.
However, not all detections are necessary in an object trajectory.
For instance, a car travels in a straight line from frame 1 to 3,
where it is at location (0, 1), (0, 2), (0, 3) at frame 1, 2, 3.
The in-between detection (0, 2) is unnecessary as the detections at frame 1 and frame 3 produce the same trajectory.

Spatialyze's {\em\sample} utilizes users' geospatial metadata ({\em\sCamera} and {\em\sRoadNetwork}) to heuristically prune out video frames that are likely to contain such unnecessary detections, 
by sampling only the frames that may contain at least one necessary detection.
As a result, the object tracker only needs to perform data association on fewer frames,
significantly reducing its runtime.

\subsubsection{High-Level Concepts}
Suppose a car is visible to a {\em\sCamera} and driving in a lane.
As discussed in \autoref{sec:cdl-world},
a {\em\cObject} of type ``car'' has 2 {inherited physical behaviors}:
it follows the lane's direction and travels at the speed limit regulated by traffic rules.
At a particular frame,
the {\em\sample} leverages these behaviors to estimate the trajectory of the car so that the object tracker need not
track any frame until:
i) a car exits its current lane;
ii) a car exits the {\em\sCamera}'s view;
iii) a new car enters the {\em\sCamera}'s view.
These events are {\em sampleEvents}.
The {\em\sample} samples the frame when a {\em sampleEvent} happens for the object tracker to
track.

For (i),
when the car is in a lane, without object tracking,
the {\em\sample} assumes that it follows the lane's direction until it reaches the end of the lane.
After it exits the lane, it may enter an intersection and stop traveling in a straight line.
To maintain the accuracy of the trajectory,
the object tracker needs to perform data association at the frame right before the car exits the lane.
In contrast, if the car is already at an intersection,
it may not travel in a straight line;
therefore, the object tracker cannot skip any frame.

Event (ii) is
the first frame that the car becomes invisible to the {\em\sCamera}.
If this event happens before the car exits its lane,
although its trajectory continues along the lane,
the object tracker needs to perform data association at the frame before the car becomes invisible to the {\em\sCamera} as the end of the car's trajectory.

For (iii),
if a new car enters the camera's view,
the object tracker needs to start tracking the new car at that frame.

If a frame contains multiple cars, the {\em\sample} skips to the earliest frame that
(i), (ii), or (iii)
is estimated to happen for any car,
which prevents incorrect tracking for any of them.
Our sampling algorithm implements this idea.
Given a current frame and its detected cars,
it
samples the next frame for the object tracker to track.
This algorithm starts from the first frame of a video to find the next frame, skips to it, and repeats until the end of the video.

Because the algorithm depends on the 2 mentioned {inherited physical behaviors},
it only works when users filter for {\em\cObjects} with such behaviors,
such as vehicles.
Therefore, the video processor only adds the {\em\sample} when users' workflow {\em Filters} on the object type being vehicles.

\subsubsection{Sampling Algorithm}
\label{sec:sample-algo}
\autoref{listing:se} shows the pseudocode of each {\em sampleEvent}.
\texttt{\textbf{\small exitsLane}} implements (i) when the car exits the lane.
Given the car's location at the current frame ({\small \tt carLoc}) and the lane that contains the car,
we assume that the car drives in the lane straight along the lane's direction.
In line~\ref{line:exit-lane-loc}, we model the car's movement as a motion tuple ({\small \tt carLoc}, {\small \tt lane.direction}).
We compute the intersection between the car's motion tuple and the lane as the location where the car exits the lane.
Given the car's speed {\small \tt v}, we know the time it reaches the exit location starting from the current frame in line~\ref{line:exit-lane-time}.
We sample the frame before the car reaches the exit location.
\texttt{\textbf{\small exitsCamera}} implements (ii) when the car exits the camera's view.
Knowing the car's current location, moving direction, and speed,
{\small \tt carMoves} computes its location at any given frame in line~\ref{line:car-moves}.
In line~\ref{line:view-contain}, we use the viewable area of the {\em\sCamera} at each frame derived from \autoref{sec:inview} to find the first frame when the car's location is not in the camera's view.
This is the frame when the car already exits the camera's view,
so we sample its preceding frame.
\texttt{\textbf{\small newCar}} implements (iii) when a new car enters the {\em\sCamera}'s view.
In line~\ref{line:more-car}, we sample the next frame that has more cars detected than the current one.
The number of cars is a good heuristic to predict that a new car is entering the frame.

At the current frame,
the {\em\sample} first finds the frame in which we have more detections for {\em sampleEvent} (iii). 
For each car, the {\em\sample} samples frames for {\em sampleEvents} (i) and (ii).
Among (i), (ii), (iii), the {\em\sample} choose the earliest {\em sampleEvent}
as the next frame for the object tracker to track.

\vspace{2pt}
\aboveCodeVSpace%
\begin{lstlisting}[language=Python, mathescape=true, escapeinside=!!, caption=\textmd{Computing the events that might trigger a frame sample}, label=listing:se]
def exitsLane(currentFrame, carLoc, lane, v, cameras):
    exitLaneLoc = intersection((carLoc, lane.direction), lane)!$\label{line:exit-lane-loc}$!
    exitLaneTime = cameras[currentFrame].time + distance(exitLaneLoc, carLoc)/v!$\label{line:exit-lane-time}$!
    return lastFrameBefore(exitLaneTime) !$\label{line:last-frame}$!
def exitsCamera(currentFrame, carLoc, lane, v, cameras):
    nextFrame = currentFrame + 1
    while cameras[nextFrame].view.contains(carMoves(carLoc, lane.direction, v,!$\label{line:view-contain}$!
          cameras[nextFrame].time - cameras[currentFrame].time)): nextFrame += 1!$\label{line:car-moves}$!
    return lastFrameBefore(nextFrame)
def newCar(currentFrame):
    nextFrame = currentFrame + 1
    while numCars(nextFrame) <= numCars(currentFrame): nextFrame += 1 !$\label{line:more-car}$!
    return nextFrame
\end{lstlisting}%
\belowCodeVSpace

\paragraph{Sampling Algorithm Walk-through}
Frame number 1 labeled as {\em currentFrame} in \autoref{fig:skipping} was taken at 3:51:49.5 PM.
Now, we illustrate how the {\em\sample} gets the next frame to sample.
In this frame, our object detector only detects one car, $Car_A$.
The 3D Location Estimator step calculates its 3D location to be (70.92, 74.7, 0).
We join the car's location with the polygons of traffic lanes in our geospatial metadata store to get the geospatial information of traffic lane $Lane_A$ that contains the car, specifically, the polygon of $Lane_A$ and its direction to be 181\textdegree~counterclockwise from the east. We speed up this process by using the geospatial index for the polygons.
We assume the car's speed is 25 mph based on common traffic rules.
We also have the viewable area %
of all frames.

First, we compute {\small \tt \textbf{newCar}(1)} when new cars enter $Camera_B$'s view.
In the rest of the videos, only one car is in the frame.
We then compute {\small \tt \textbf{exitsLane}(1, \allowbreak(70.92,74.7,0), \allowbreak $Lane_A$, \allowbreak 
25 mph, \allowbreak $cameras$)} when $Car_A$ exits $Lane_A$.
In line~\ref{line:exit-lane-loc}, $Car_A$'s motion tuple intersects the polygon of $Lane_A$ at location (66.3, 72.7, 0)
at 3:51:49.96 PM in line~\ref{line:exit-lane-time}.
The last frame before this time is frame \bm{22},
depicted as {\small \tt exitsLane} in \autoref{fig:sample-plans}.
Finally, we compute {\small \tt \textbf{exitsCamera}(1, \allowbreak
(70.92,74.7,0), \allowbreak $Lane_A$, \allowbreak 25 mph, \allowbreak $cameras$)}
when the $Car_A$ exits the $Camera_B$'s view.
The $Car_A$ exits the camera view at frame \bm{35} computed in line~\ref{line:car-moves}.
Thus,
we sample the frame \bm{34}, 
depicted as {\small \tt exitsCamera} in \autoref{fig:sample-plans}.

Among the 3 {\em sampleEvents}, {\em exitsLane} in \autoref{fig:sample-plans}
is the earliest at frame \bm{22}.
Thus, the {\em\sample} samples the frame before $Car_A$ exits its lane at frame \bm{22} labeled as {\em nextFrame} in \autoref{fig:skipping},
skipping all frames in between.
The {\em\sample} repeats the algorithm to sample frames until it reaches the end of the video.

\subsubsection{Benefits and Limitations}
\label{sec:sample-limit}
\begin{figure}%
  \centering%
  \includegraphics[width=\columnwidth]{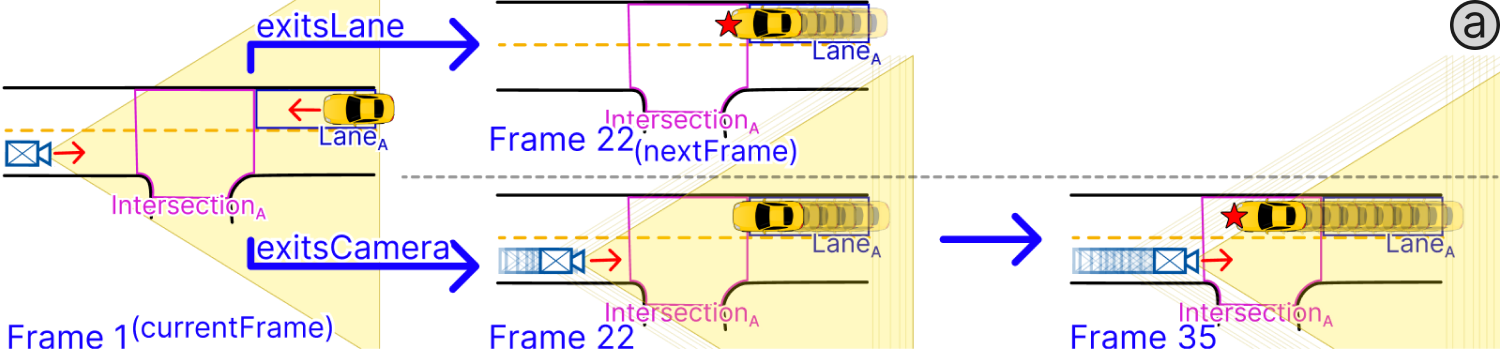}%
  \vspace{2pt}
  \includegraphics[width=\columnwidth]{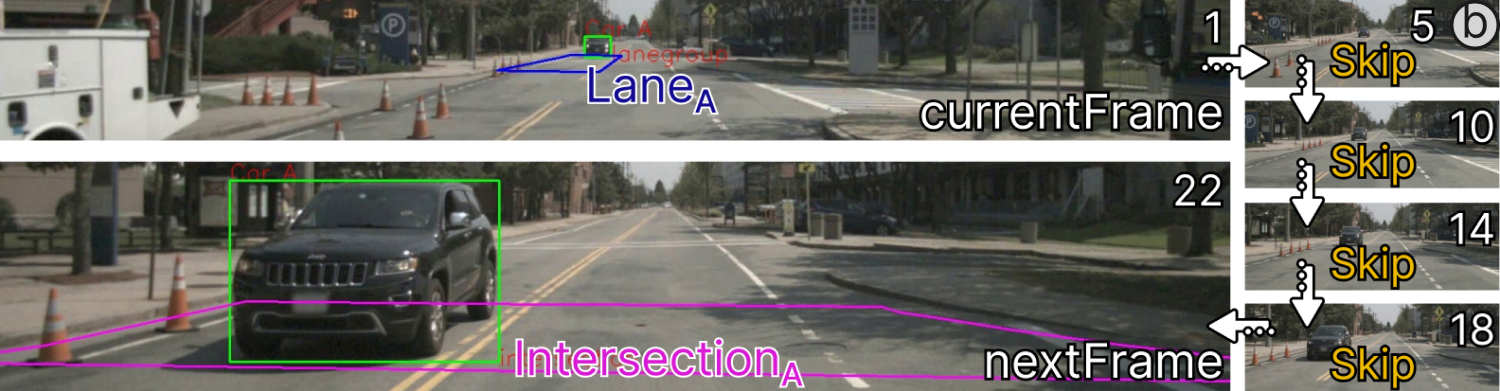}%
  \vspace{4pt}
  \includegraphics[width=\columnwidth]{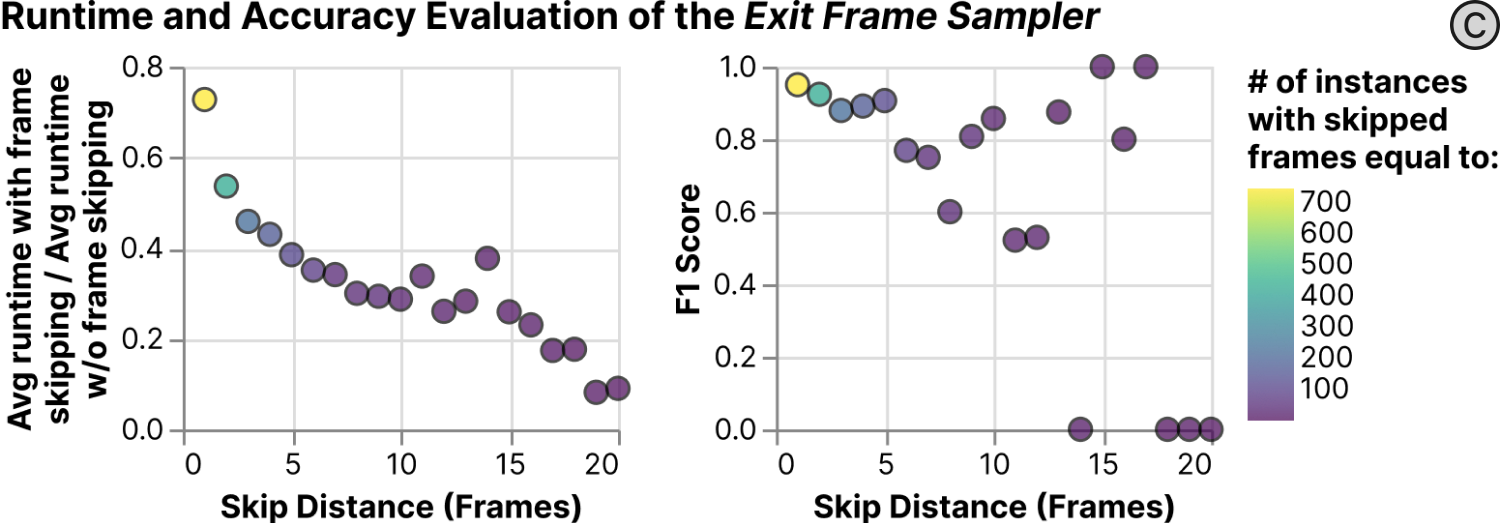}%
  \Description{%
  The figure above is an illustration of exitLane event and exitCamera event.
  The exitLane event happens earlier.
  In the figure below, the image shows the first frame that the sampling algorithm starts running on,
  the frames that the algorithm skips,
  and the frame that the algorithm skips to
  (when the car in the first frame exits the lane that it was on).
  Runtime and Accuracy Evaluation of the Exit Frame Sampler. The left scatter plot shows the object tracking time reduces when the Skip Distance is higher. The right scatter plot shows that the F1 Score (accuracy) of the object tracking algorithm decreases when the Skip Distance is higher. The plot shows our observation that there are more instances of small skipping as a result of the sampling algorithm than instances of large skipping.
  }%
  \caption{\bigcircled{a} Sample Events exitsLane (i) and exitsCamera (ii).
  \bigcircled{b} Example Results of our Sampling Algorithm.
  \bigcircled{c} F1 score and runtime reduction for each skip distance when the
  {\em\sample} can skip at least 1 frame.%
  }\belowCapVSpace%
  \label{fig:skipping}%
  \label{fig:sample-plans}%
  \label{fig:exit-frame-sampler-benchmark}%
\end{figure}

To evaluate the efficacy of the {\em\sample}, we define ``skip distance'' as the number of frames skipped between two non-skipped frames.
We quantify tracking performance using the {\em\sample} by summing the runtime of the sampling algorithm and StrongSORT from the current frame to the next non-skipped frame and dividing that by the runtime without running the {\em\sample}.
After processing all the videos from the nuScenes dataset, we computed the average for all data points with the same skip distance.
As \autoref{fig:exit-frame-sampler-benchmark} shows,
the runtime ratio decreases as the skip distance increases.
With an average skip of \bmv{\evalVariableOneDeQueryTwoSkipDistance} frames,
per-frame runtime is \bmv{\evalVariableOneDeAllSSReductionDistance\%} of the original.

Meanwhile, skipping too many frames results in the tracker losing sight of the target vehicles.
To assess this risk,
we computed the F1 score at each skip distance.
The ground truth is the object trackings using StrongSORT without the {\em\sample}.
Here, ``prediction'' refers to whether an object at frame $f$ is correctly tracked at frame $(f + \textrm{skip distance})$ as predicted by the {\em\sample}.
\autoref{fig:exit-frame-sampler-benchmark} shows that the accuracy stays high for skip distances
at 13
with the runtime of
\bmv{\evalVariableReductionTwentyDistance\%} of the original per frame.
Hence, we set the maximum skip distance to 13 to maximize this accuracy vs runtime tradeoff.
We omit the skip distances over 20 from the graph due to their poor accuracy
to focus on effective metrics.%

\section{Evaluation}
\label{sec:evaluation}

We have implemented a prototype of Spatialyze and evaluated its components against other
SOTA
systems.
We also evaluate our optimization techniques with an ablation study
by comparing Spatialyze's video processing plans with and without each of them.
We explain our experiment setup below.

{\em Queries.}
We implemented Spatialyze's workflows using four realistic road scenarios,
inspired by queries from Scenic~\cite{kim:scenic-val}, VIVA~\cite{romero:viva}, and SkyQuery~\cite{bastani:skyquery}. 
These workflows test Spatialyze against real-life situations.
Each workflow starts by loading \emph{\sRoadNetworks} and \emph{\sVideos} with their \emph{\sCameras} from each query's corresponding datasets,
similar to lines~\ref{line:world}-\ref{line:end-build} in \autoref{listing:python-workflow}.
Then, we filter the videos with one of the queries described in \autoref{tab:queries},
similar to lines~\ref{line:object}-\ref{line:pred-heading}.
All queries look for objects closer than 50 meters.
Finally, we save the filtered videos into video snippet files, similar to line~\ref{line:observe}.
\autoref{tab:queries} shows the queries in natural language and S-Flow's filter predicates.
\begin{table}[!ht]%
\caption{Evaluation query \hl{descriptions} and {their {S-Flow} {predicates}}
(VisualRoad$^\text{\scriptsize [vr]}$; Scenic$^\text{\scriptsize [s]}$;
VIVA$^\text{\scriptsize [vv]}$; and SkyQuery$^\text{\scriptsize [vv]}$).}%
\tableAboveVSpace%
\label{tab:queries}%
\scriptsize
\begin{tabular}{|p{.15cm}|p{7.6cm}|}
    \hline
    \hspace*{-.15cm}
    {\bf Q1} &  
    \hspace*{-.1cm}
    \hl{A \emph{pedestrian} at an intersection facing perpendicularly to that of a \emph{camera} (vr, s).}\newline
    {\texttt{\textbf{\tiny contains(intersection,person) \& perpendicular(person,camera)}}}\\
    \hline
    \hspace*{-.15cm}
    {\bf Q2} &  
    \hspace*{-.1cm}
    \hl{\emph{2 cars} at an intersection moving in opposite directions (vr, s).}\newline
    {\texttt{\textbf{\tiny contains(intersection,[car1,car2]) \& opposite(car1,car2)}}}\\
    \hline
    \hspace*{-.15cm}
    {\bf Q3} &
    \hspace*{-.1cm}
    \hl{\emph{Camera} moving in the direction opposite to the lane direction,
    with another \emph{car} (which is moving in the direction of the lane) within 10 meters of it (vr, s).}
    {\texttt{\textbf{\tiny contains(lane,[camera, car]) \& opposite(lane,camera) \& sameDirection(lane,car) \& distance(camera,car)<10}}}\\
    \hline
    \hspace*{-.15cm}
    {\bf Q4} &
    \hspace*{-.1cm}
    \hl{A \emph{car} and a \emph{camera} moving in the same direction on lanes;
    \emph{2 other cars} moving together on opposite lanes (vr, s).}
    {\texttt{\textbf{\tiny contains(lane1,[car1,camera]) \& sameDirection(car1,camera) \& contains(lane2,[car2,car3]) \& sameDirection(car2,car3) \& opposite(lane1,lane2)}}}\\
    \hline
    \hspace*{-.15cm}
    {\bf Q5} &
    \hspace*{-.1cm}
    \hl{A \emph{pedestrian} is at an intersection (vr, s).}
    {\texttt{\textbf{\tiny contains(intersection,person)}}}\\
    \hline
    \hspace*{-.15cm}
    {\bf Q6} &
    \hspace*{-.1cm}
    \hl{\emph{2 cars} are at an intersection (vr, s).}
    {\texttt{\textbf{\tiny contains(intersection,[car1,car2])}}}\\
    \hline
    \hspace*{-.15cm}
    {\bf Q7} &
    \hspace*{-.1cm}
    \hl{A \emph{car} is on a lane within 10 meters of the \emph{camera} (vr, s).}\newline
    {\texttt{\textbf{\tiny contains(lane,camera) \& distance(camera,car)<10}}}\\
    \hline
    \hspace*{-.15cm}
    {\bf Q8} &
    \hspace*{-.1cm}
    \hl{\emph{3 cars}, each on a lane (vr,s).}
    {\texttt{\textbf{\tiny contains(lane,car1) \& contains(lane,car2) \& contains(lane,car2)}}}\\
    \hline
    \hspace*{-.15cm}
    {\bf Q9} &
    \hspace*{-.1cm}
    \hl{A \emph{car} turning left with a \emph{pedestrian} at an intersection (vr, vv).}\newline
    {\texttt{\textbf{\tiny contains(intersection,[car,person]) \& turnLeft(car)}}}\\
    \hline
    \hspace*{-.15cm}
    {\bf Q10} &
    \hspace*{-.1cm}
    \hl{A \emph{car} stopped in a cycling lane (vr, sq).}
    {\texttt{\textbf{\tiny contains(bikeLane,car) \& stopped(car)}}}\\
    \hline
\end{tabular}%
\tableBelowVSpace%
\end{table}%

{\em Dataset.}
We use the multi-modal AV datasets from Scenic~\cite{kim:scenic-val,fremont:scenic-data,fremont:scenic-scene} to evaluate Spatialyze.
They comprise 3 datasets that are needed for executing the 4 queries mentioned:
1) videos from on-vehicle cameras,
2) the cameras' movement and specification,
and 3) the road network of Boston Seaport where the videos were shot.
The former 2 datasets are directly from nuScenes's training set~\cite{caesar:nuscenes},
and the latter is from Scenic.
From nuScenes, we randomly sampled \bmv{\NumScenes} out of 467 scenes that are captured at the Boston Seaport from on-vehicle cameras;
we chose videos from 3 front cameras for each scene;
each video is 20 seconds long at 12 FPS.
The runtime of Spatialyze's execution per video does not depend on the number of input videos;
therefore, we have determined that \bmv{\NumVideos} sampled videos sufficiently demonstrate our contributions.
To compare Spatialyze with VIVA, we use the VIVA-provided Jackson Square Traffic Camera dataset, which contains 19,469 5-second-long 1080p videos at 30 FPS of a traffic intersection, in addition to the nuScenes dataset.
Finally, to compare with SkyQuery, we use their aerial drone geospatial video datasets, which contain 17,853 frames of a 1080p top-down video, along with the per-frame GPS information.

{\em Hardware.}
In our experiments, we use Google Cloud Compute Engine with Nvidia T4;
n1-highmem-4 Machine with 2 vCPUs (2 cores), 26 GB of Memory; and
Balanced persistent disk.

\subsection{Comparison with Other Systems}
\label{sec:eval-others}
\paragraph{Sytems selection}
Since we are unaware of any end-to-end system that focuses on geospatial video analytics,
we chose 5 systems to evaluate Spatialyze, where
we selected each system to evaluate the novelty of different components of Spatialyze.

    \vspace{2.5pt}
    \noindent
    {\em\bf EVA~\cite{xu:eva}.}
    We evaluate Spatialyze against EVA
    with its SOTA optimization techniques to speed up video processing.
    
    \vspace{2.5pt}
    \noindent
    {\em\bf VIVA~\cite{romero:viva}.}
    We evaluate Spatialyze against VIVA to emphasize the novelty of our optimization techniques to speed up our
    end-to-end geospatial video query processing.
    In addition, the videos, with which VIVA uses to evaluate, are long;
    we use this fact to evaluate Spatialyze's capability of processing long videos.
    Finally, VIVA uses a different ML function~\cite{wojke:deepsort} to track objects;
    we evaluate the video processor's capability of supporting the function.
    
    \vspace{2.5pt}
    \noindent
    {\em\bf NuScenes Devkit~\cite{caesar:nuscenes}.}
    As the official tool to process NuScenes' geospatial movement data,
    we evaluate the runtime efficiency and the capability of processing large amounts of movement objects of 
    the \cObjects Query Engine against the NuScenes Devkit.
    
    \vspace{2.5pt}
    \noindent
    {\em\bf OTIF~\cite{bastani:otif}.}
    We evaluate Spatialyze against OTIF to emphasize the novelty of our optimization techniques to speed up object tracking.
    
    \vspace{2.5pt}
    \noindent
    {\em\bf SkyQuery~\cite{bastani:skyquery}.}
    We evaluate Spatialyze against SkyQuery's geospatial video query processor.
    As SkyQuery uses
    different object detector, tracker, and 3d-estimator,
    we evaluate the video processor's ability to support different ML functions
    and how the existing optimization techniques perform with the new ML functions.

\subsubsection{EVA}
\label{sec:eva}
We compare the runtime of the Spatialyze pipeline with that of EVA's implementation using the same queries.
EVA is a VDBMS that allows custom user-defined functions (UDFs) along with their runtime optimizations.
Since EVA executes the queries on a frame-by-frame basis,
it cannot compute object tracks (and hence object directions),
so we use EVA to evaluate only the object detection capability of Spatialyze with Q5-8.
To use EVA's caching mechanism,
we execute the queries on EVA when they are run in series without resetting EVA and its database (running Q6 directly after Q5, Q7 directly after Q6, etc).
The results are shown in \autoref{fig:runtime-others}.
Our goal in comparing against EVA is to evaluate the efficiency of the two system's query optimization techniques in \autoref{sec:inview} and \autoref{sec:depth}.

As shown, Spatialyze performs \bmv{2-7.3$\times$} faster than EVA on Q5-7,
since Spatialyze does not need to run Monodepth2 on each frame,
due to the {\em \objtype} and {\em \depthEstimation}.
In addition, Spatialyze shaves even more time by employing the {\em \inview} to avoid costly YOLO on frames where they are unnecessary.
For Q8, our processing time is comparable to EVA.
This is because Spatialyze's \cObjects Query Engine finds every possible triplet of cars;
thus it performs 2 self-joins internally,
while EVA only returns video frames with 3 cars or more.
We also compare the lines of code (LoC) counts between Spatialyze (14-30) and EVA (58-60),
showing that Spatialyze makes query implementation easier and requires less LoC for the same results.

\subsubsection{VIVA}
\label{sec:viva}
We compare the run-time of the Spatialyze workflow with that of VIVA using the traffic application query from VIVA's evaluation (Q9 in \autoref{tab:queries})
as it is VIVA's only query that involves geospatial content and road network. 
We run this query against both the VIVA-provided Jackson Square dataset as well as the nuScenes dataset.
To match VIVA's algorithm, we also resize Spatialyze's input video to a resolution of $360\times240$,
resample the video at 1 FPS,
and employ DeepSORT~\cite{wojke:deepsort} for tracking rather than StrongSORT to demonstrate
Spatialyze's extensibility.
We notice that Spatialyze can execute the same workflow at the videos' native resolution and FPS while VIVA crashes, which demonstrates Spatialyze's ability to process large videos
and the runtime efficiency of the video processor as a result of Spatialyze's optimization.

Shown in \autoref{fig:runtime-others}, Spatialyze performs \bmv{1.68$\times$} faster compared to VIVA when run on VIVA's dataset, and \bmv{6$\times$} faster when run on the nuScenes dataset. 
We attribute this increase in performance to Spatialyze's {\em \objtype} which eliminates the amount of time spent tracking unnecessary objects. 
VIVA also spends significantly more time creating an optimization plan.%

In addition, we compare the LoC of the implementations of VIVA (56) and Spatialyze (51), with Spatialyze providing a shorter implementation.
While we do not have the ground truth to compare with, Spatialyze has higher accuracy in the query results: 45\% of tracks returned by Spatialyze are indeed cars turning left.
Meanwhile, only 35.7\% of the tracks returned by VIVA are correct.

\subsubsection{nuScenes}
\label{sec:devkit}
We compare the run-time of the \emph{\cObjects Query Engine} stage with the same queries implemented using the nuScenes Devkit~\cite{caesar:nuscenes}. 
NuScenes Devkit is a Python API used to query results from the nuScenes dataset~\cite{caesar:nuscenes}.
We only compare the {\cObjects Query Engine} stage as Devkit operates on already processed videos and ingested object annotations.

The Devkit implementation of Q4 ran out of memory during execution due to the creation of too many Python objects,
as a result of materializing all possible combinations of 3 object tuples in memory before filtering them.
Shown in \autoref{fig:runtime-others},
it also takes significantly longer to execute the same query as Spatialyze,
with Spatialyze achieving a \bm{117-716$\times$} faster runtime. 
Two of the main factors that allow Spatialyze to achieve this better execution time are the following.
First, Spatialyze pre-processes the road network data, generating additional columns and indexes that allow it to avoid costly joins, which contribute greatly to the large execution time of Devkit.
Second, certain Devkit functions perform costly linear algebra that is avoided by Spatialyze using the geospatial metadata store. 
Comparing the line counts between the different systems,
it took 55-123 lines in Devkit to implement the queries while it only took
12-29 lines to do so in S-Flow.

\subsubsection{OTIF}
\label{sec:otif}
We compare the run-time of the video processor of Spatialyze with that of OTIF on object tracking.
We run each of the two systems on the sampled nuScenes datasets
and compare the number of frames processed per second for each system.
OTIF's training is \bmv{61m37s}.
After it is trained,
OTIF tracks objects at \bmv{17.3} FPS. %
For Spatialyze, our video processor applies all optimizations elaborated in \autoref{sec:optimization},
and it processes videos at \bm{18.3-39.5} FPS to get the tracks for each query.
Without counting OTIF's training time, Spatialyze still tracks objects faster.%

\subsubsection{SkyQuery}
\label{sec:skyquery}
SkyQuery is an aerial drone sensing platform for detecting and tracking vehicles from top-down videos.
A user inputs an aerial drone's video and its geospatial data, then constructs a query to detect and track objects.
Spatialyze and SkyQuery both process geospatial videos by
1) detecting objects of interest, 2) finding the objects' 3D locations, and then 3) tracking the objects.

\begin{figure}%
    \centering%
    \includegraphics[width=\columnwidth]{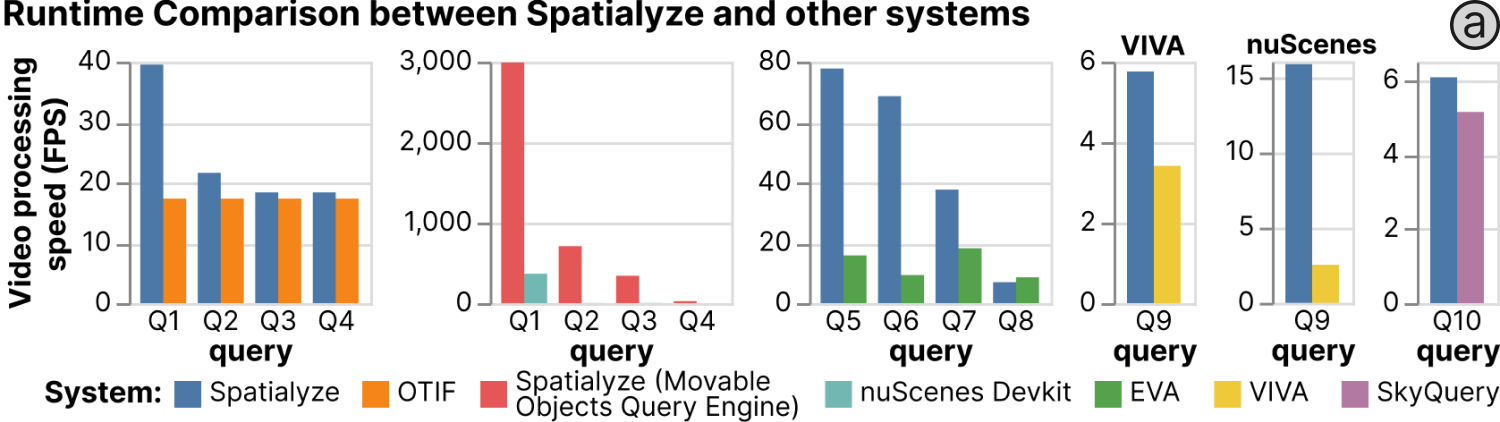}%
  \vspace{4pt}
    \includegraphics[width=\columnwidth]{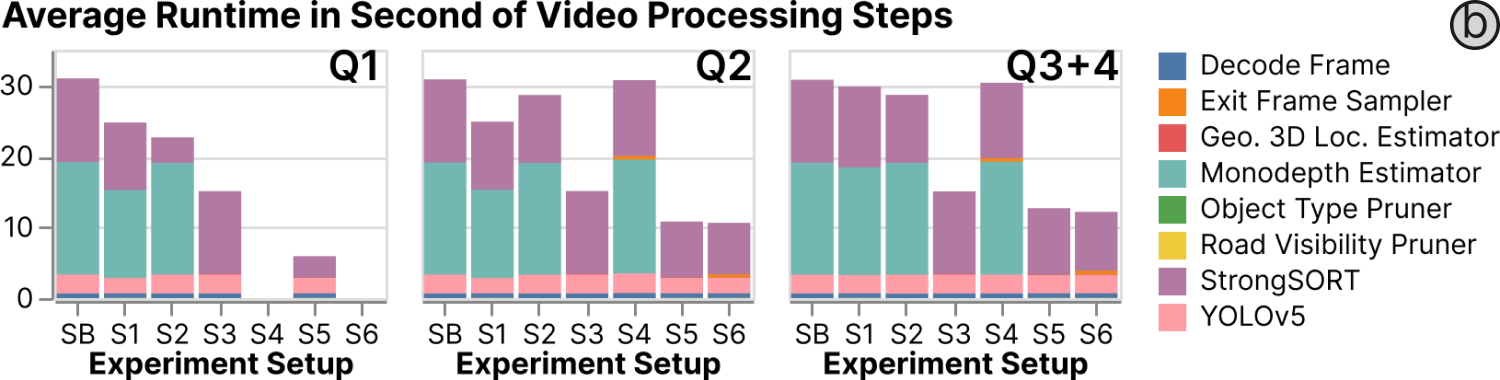}%
  \vspace{4pt}
    \includegraphics[width=\columnwidth]{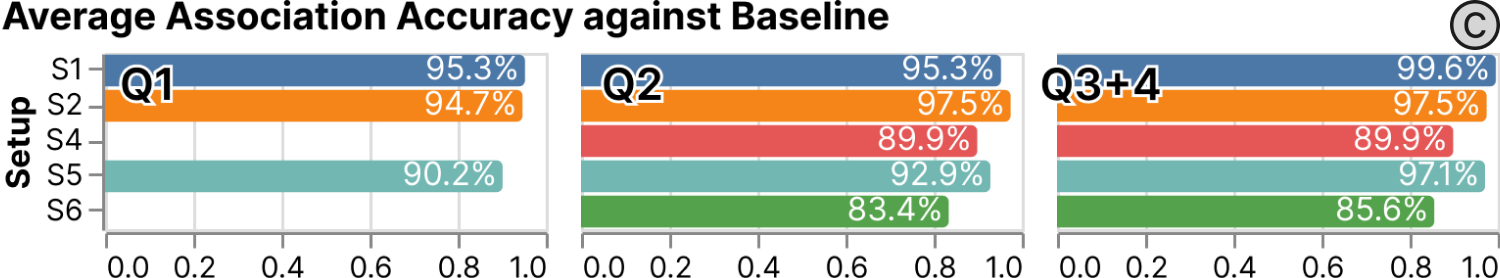}%
    \aboveCapVSpaceTwo%
    \Description{%
    Runtime comparison of Spatialyze against OTIF, NuScenes Devkit, EVA, VIVA (both their data and NuScenes), and SkyQuery (with their data).
    Spatialyze is faster in all cases except when running Q8 compared to EVA
    
    The bar chart above shows the runtime breakdown of each streaming operator, comparing between SB, S1, S2, S3, S4, S5, S6, and for Q1, Q2, Q3, Q4. The chart below shows the AssA against SB of S1, S2, S4, S5, S6, and for Q1, Q2, Q3, Q4.}%
    \caption{%
    \bigcircled{a} Video frames processed per second for each system.
    \bigcircled{b} Average video processing time for each video of 20 seconds, comparing
    each optimization technique.
    \bigcircled{c} AssA against (SB) vs. other experiment setups.}\belowCapVSpace%
    \label{fig:runtime-others}%
    \label{fig:runtime-ablation}%
\end{figure}

To compare with SkyQuery,
we implemented the 3 steps mentioned as Spatialyze's video processing steps,
and used SkyQuery's customized YOLOv3 for car detection, their 3D location estimator, and SORT~\cite{bewley:sort} for object tracking.
We retained the rest of Spatialyze's components as well as the optimization steps.
Then, we executed both of the systems on Q10 on SkyQuery's video and geospatial metadata, and compared execution times.
As shown in \autoref{fig:runtime-others}, Spatialyze processes 6.08 FPS, while SkyQuery processes 5.15 FPS, making us 18\% faster.
We write 15 lines of S-Flow vs 4 lines of SkyQuery to express Q10.
The majority of the lines in S-Flow are spent on setting up the world and inputting
videos and geospatial metadata.
While we input videos and geospatial metadata into Spatialyze using S-Flow,
we need to specify such input directories in the command line when starting SkyQuery.

As our evaluation uses the same ML models as SkyQuery, our speedup is entirely due to Spatialyze's leveraging of the query's semantics to prune video frames and avoid executing ML functions on them.
Specifically, Spatialyze uses the {\em\inview} to exclude video frames that do not contain any cycling lanes.
We do not evaluate our accuracy in this evaluation as we use the same ML inference functions as those in SkyQuery.
As the only optimization technique applied in this evaluation is the {\em\inview},
it only prunes out video frames that do not contain a cycling lane, and thus, does not affect the query result.

\subsection{Ablation Study}
\label{sec:ablation}
To evaluate each of our optimization techniques,
we manually construct video processing plans for the following 7 experiment setups.
\textbf{(SB) Baseline} is a Baseline plan without any optimization.
It includes the 4 processing operators discussed in \autoref{sec:video-processing}.
\textbf{(S1)} is the Baseline plan with the \emph{\inview}.
\textbf{(S2)} is the Baseline plan with the \emph{\objtype}.
\textbf{(S3)} is the Baseline plan with the Monocular Depth Estimator replaced with the \emph{\depthEstimation}.
\textbf{(S4)} is the Baseline plan with the \emph{\sample}.
\textbf{(S5)} enables all optimizations except the \emph{\sample}.
\textbf{(S6)} enables all optimizations.
We evaluated Spatialyze using Q1-4 because Spatialyze can apply
all or most of the optimization techniques for the execution of such queries;
therefore, we can compare the effects of each optimization technique.

\subsubsection{Execution time}
\label{sec:ablation-time}
The average runtime of our workflow execution without any optimization is \bmv{\evalVariableZeroOverallRuntime} seconds for each video.
When broken down, \bmv{\evalVariableTwoIngestPercent\%} of the time is spent in the {\em Data Integrator};
\bmv{\evalVariableOneVProcessPercent\%} in {\em Video Processor};
\bmv{\evalVariableOneQueryPercent\%} in {\em \cObjects Query Engine};
\bmv{\evalVariableOneSavePercent\%} in {\em Output Composer}.
In this section, we evaluate the effect of each of our optimization techniques,
by executing Spatialyze on the Q1-Q4.
Because the video processing plans for Q3 and Q4 are the same, we do not discuss Q4 in this section.

With all optimization techniques,
we achieve \bmv{\evalVariableOneOverallQueryThreeFourRuntimeSpeedup-\evalVariableOneOverallQueryOneRuntimeSpeedup$\times$} speed up over the baseline plan,
as shown in \autoref{fig:runtime-ablation}.
The \emph{\inview} affects the entire plan.
It prunes out \bmv{\evalVariableOneInviewQueryThreeFourSkip\%} of the frames that do not contain ``lane'' in Q3 and \bmv{\evalVariableOneInviewQueryOneSkip\%} of the frames that do not contain ``intersection'' in Q1-2,
with \bmv{\evalVariableOneInviewQueryThreeFourRuntimeReduction-\evalVariableOneInviewQueryOneTwoRuntimeReduction\%} runtime reduction.
The \emph{\objtype} only affects StrongSORT.
It prunes out \bmv{\evalVariableOneObjTypeQueryTwoObjReduction\%} of objects that are not cars in Q2-4 or trucks and \bmv{\evalVariableOneObjTypeQueryOneObjReduction\%} that are not pedestrians in Q1,
with \bmv{\evalVariableOneObjTypeQueryTwoThreeFourRuntimeReduction-\evalVariableOneObjTypeQueryOneRuntimeReduction\%} runtime reduction.
The \emph{\depthEstimation} makes the runtime portion of 3D location estimation insignificant (from \bmv{\evalVariableZeroMonodepthAllPercent\%} to \bmv{\evalVariableTwoDepthAllTDEstmPortionPercent\%}).
Q1 does not execute the \emph{\sample} as it only works with cars or trucks.
The \emph{\sample} has an overhead runtime of executing \autoref{sec:sample-algo};
however, it reduces the number of frames into StrongSORT (\bmv{\evalVariableOneDeQueryTwoSkipPercent\%}),
with \bmv{\evalVariableOneDeAllRuntimeReductionPercent\%} runtime reduction.
We also evaluate the execution time with respect to video length by varying the video length input into the workflow.
We found that the execution time of the workflow is linearly proportional to video length and Spatialyze can process videos larger than the available memory.

\subsubsection{Accuracy}
\label{sec:accuracy}
Our optimization techniques potentially reduce object tracking accuracy as a result of dropping frames.
To quantify this, 
we evaluate our optimization techniques using the execution results from \textbf{(SB)} setup as the ground truth.
Then, we compare the execution results from other setups against that of the \textbf{(SB)}.

We evaluate our tracking accuracy of each experiment setup by comparing the tracking output from the experiment setup with the tracking output from \textbf{(SB)} setup, using HOTA~\cite{luiten:hota},
an evaluation metric for measuring multi-object tracking accuracy.
Using the ground truth tracks from a video and prediction tracks from the same video,
HOTA returns an Association Accuracy (AssA) score.
In this evaluation, the ground truth tracks are the tracking results from the \textbf{(SB)} setup,
and the prediction tracks are the tracking results from each of the other setups.
We also exclude object detections from the frame pruned by the \emph{\inview} because this pruning is a part of users' predicates and will not be output.%

In \autoref{fig:runtime-ablation}, the {\em \inview} {\bf (S1)} causes a small AssA drop of \bmv{\evalVariableOneQueryThreeFourInViewAccuracyDrop\%} for Q3-4 and \bmv{\evalVariableOneQueryOneInViewAccuracyDrop\%} for Q1-2.
The {\em \objtype} {\bf (S2)} also causes a small AssA drop of \bmv{\evalVariableOneQueryTwoObjectFilterAccuracyDrop\%} from Q2-4
and \bmv{\evalVariableOneQueryOneObjectFilterAccuracyDrop\%} for Q1.
We do not include {\bf (S3)} into the results since the object tracker only concerns the 2D bounding boxes of objects;
{\em \depthEstimation} does not affect the AssA of the object tracker.
With all 3 optimization techniques {\bf (S5)},
the video processor still produces accurate object tracks of \bmv{\evalVariableOneAllOptimizedAccuracy\%} on average compared to {\bf (SB)}, while speeding up \bmv{\evalVariableOneOverallQueryThreeFourRuntimeSpeedup-\evalVariableOneOverallQueryOneRuntimeSpeedup$\times$} of its runtime.
With all optimization techniques {\bf (S6)}, the video processor's accuracy for object tracking is \bmv{\evalVariableOneAllOptimizedDEAccuracy\%} on average,
while speeding up \bmv{\evalVariableTwoOverallDeQueryTwoAdditionalSpeedup-\evalVariableTwoOverallDeQueryThreeFourAdditionalSpeedup\%}, additionally.

\section{Conclusion}

We presented Spatialyze, a system for geospatial video analytics that leverages geospatial metadata and physical properties of objects in videos to speed up video processing. 
Spatialyze can process videos of arbitrary length by streaming each frame through the operators in workflows expressed using S-Flow.
Spatialyze's DSL
allows users to specify their geospatial video workflows declaratively,
and our evaluation shows Spatialyze's efficiency in processing
such
workflows when compared to SOTA systems.

\begin{acks}
We thank Daniel Kang, Francisco Romero, Favyen Bastani, Edward Kim, and our anonymous reviewers for their insightful feedback.
This work is supported in part by NSF grants
IIS-1955488,
IIS-2027575,
ARO W911NF2110339,
ONR N00014-21-1-2724, and
DOE awards DE-SC0016260 and
DE-SC0021982.
\end{acks}

\clearpage
\bibliographystyle{ACM-Reference-Format}
\begin{CJK*}{UTF8}{gbsn}
  \bibliography{main}


\begin{thebibliography}{43}

%%% the Web address.

\ifx \showCODEN    \undefined \def \showCODEN     #1{\unskip}     \fi
\ifx \showDOI      \undefined \def \showDOI       #1{#1}\fi
\ifx \showISBNx    \undefined \def \showISBNx     #1{\unskip}     \fi
\ifx \showISBNxiii \undefined \def \showISBNxiii  #1{\unskip}     \fi
\ifx \showISSN     \undefined \def \showISSN      #1{\unskip}     \fi
\ifx \showLCCN     \undefined \def \showLCCN      #1{\unskip}     \fi
\ifx \shownote     \undefined \def \shownote      #1{#1}          \fi
\ifx \showarticletitle \undefined \def \showarticletitle #1{#1}   \fi
\ifx \showURL      \undefined \def \showURL       {\relax}        \fi
% invisible to TeX
\providecommand\bibfield[2]{#2}
\providecommand\bibinfo[2]{#2}
\providecommand\natexlab[1]{#1}
\providecommand\showeprint[2][]{arXiv:#2}

\bibitem[\protect\citeauthoryear{Anwar}{Anwar}{2022}]%
        {anwar:camera}
\bibfield{author}{\bibinfo{person}{Aqeel Anwar}.}
  \bibinfo{year}{2022}\natexlab{}.
\newblock \bibinfo{title}{What are Intrinsic and Extrinsic Camera Parameters in
  Computer Vision?}
\newblock
  \bibinfo{howpublished}{\url{https://towardsdatascience.com/what-are-intrinsic-and-extrinsic-camera-parameters-in-computer-vision-7071b72fb8ec}}.
\newblock
\newblock
\shownote{Accessed: 2023-07-25.}


\bibitem[\protect\citeauthoryear{Bastani, He, Balasingam, Gopalakrishnan,
  Alizadeh, Balakrishnan, Cafarella, Kraska, and Madden}{Bastani
  et~al\mbox{.}}{2020}]%
        {bastani:miris}
\bibfield{author}{\bibinfo{person}{Favyen Bastani}, \bibinfo{person}{Songtao
  He}, \bibinfo{person}{Arjun Balasingam}, \bibinfo{person}{Karthik
  Gopalakrishnan}, \bibinfo{person}{Mohammad Alizadeh}, \bibinfo{person}{Hari
  Balakrishnan}, \bibinfo{person}{Michael Cafarella}, \bibinfo{person}{Tim
  Kraska}, {and} \bibinfo{person}{Sam Madden}.}
  \bibinfo{year}{2020}\natexlab{}.
\newblock \showarticletitle{MIRIS: Fast Object Track Queries in Video}. In
  \bibinfo{booktitle}{\emph{Proceedings of the 2020 ACM SIGMOD International
  Conference on Management of Data}} (Portland, OR, USA)
  \emph{(\bibinfo{series}{SIGMOD '20})}. \bibinfo{publisher}{Association for
  Computing Machinery}, \bibinfo{address}{New York, NY, USA},
  \bibinfo{pages}{1907–1921}.
\newblock
\showISBNx{9781450367356}
\urldef\tempurl%
\url{https://doi.org/10.1145/3318464.3389692}
\showDOI{\tempurl}


\bibitem[\protect\citeauthoryear{Bastani, He, Jiang, Bastani, and
  Madden}{Bastani et~al\mbox{.}}{2021}]%
        {bastani:skyquery}
\bibfield{author}{\bibinfo{person}{Favyen Bastani}, \bibinfo{person}{Songtao
  He}, \bibinfo{person}{Ziwen Jiang}, \bibinfo{person}{Osbert Bastani}, {and}
  \bibinfo{person}{Sam Madden}.} \bibinfo{year}{2021}\natexlab{}.
\newblock \showarticletitle{SkyQuery: An Aerial Drone Video Sensing Platform}.
  In \bibinfo{booktitle}{\emph{Proceedings of the 2021 ACM SIGPLAN
  International Symposium on New Ideas, New Paradigms, and Reflections on
  Programming and Software}} (Chicago, IL, USA) \emph{(\bibinfo{series}{Onward!
  2021})}. \bibinfo{publisher}{Association for Computing Machinery},
  \bibinfo{address}{New York, NY, USA}, \bibinfo{pages}{56–67}.
\newblock
\showISBNx{9781450391108}
\urldef\tempurl%
\url{https://doi.org/10.1145/3486607.3486750}
\showDOI{\tempurl}


\bibitem[\protect\citeauthoryear{Bastani and Madden}{Bastani and
  Madden}{2022}]%
        {bastani:otif}
\bibfield{author}{\bibinfo{person}{Favyen Bastani} {and}
  \bibinfo{person}{Samuel Madden}.} \bibinfo{year}{2022}\natexlab{}.
\newblock \showarticletitle{OTIF: Efficient Tracker Pre-Processing over Large
  Video Datasets}. In \bibinfo{booktitle}{\emph{Proceedings of the 2022
  International Conference on Management of Data}} (Philadelphia, PA, USA)
  \emph{(\bibinfo{series}{SIGMOD '22})}. \bibinfo{publisher}{Association for
  Computing Machinery}, \bibinfo{address}{New York, NY, USA},
  \bibinfo{pages}{2091–2104}.
\newblock
\showISBNx{9781450392495}
\urldef\tempurl%
\url{https://doi.org/10.1145/3514221.3517835}
\showDOI{\tempurl}


\bibitem[\protect\citeauthoryear{Bewley, Ge, Ott, Ramos, and Upcroft}{Bewley
  et~al\mbox{.}}{2016}]%
        {bewley:sort}
\bibfield{author}{\bibinfo{person}{Alex Bewley}, \bibinfo{person}{Zongyuan Ge},
  \bibinfo{person}{Lionel Ott}, \bibinfo{person}{Fabio Ramos}, {and}
  \bibinfo{person}{Ben Upcroft}.} \bibinfo{year}{2016}\natexlab{}.
\newblock \showarticletitle{Simple online and realtime tracking}. In
  \bibinfo{booktitle}{\emph{2016 IEEE International Conference on Image
  Processing (ICIP)}}. \bibinfo{pages}{3464--3468}.
\newblock
\showISSN{2381-8549}
\urldef\tempurl%
\url{https://doi.org/10.1109/ICIP.2016.7533003}
\showDOI{\tempurl}


\bibitem[\protect\citeauthoryear{Bradski}{Bradski}{2000}]%
        {bradski:opencv}
\bibfield{author}{\bibinfo{person}{Gary Bradski}.}
  \bibinfo{year}{2000}\natexlab{}.
\newblock \showarticletitle{The {OpenCV} Library}.
\newblock \bibinfo{journal}{\emph{Dr. Dobb's Journal of Software Tools}}
  \bibinfo{volume}{25}, \bibinfo{number}{11} (\bibinfo{date}{Nov.}
  \bibinfo{year}{2000}), \bibinfo{pages}{120, 122--125}.
\newblock
\showCODEN{DDJOEB}
\showISSN{1044-789X}
\urldef\tempurl%
\url{http://www.ddj.com/ftp/2000/2000_11/opencv.txt}
\showURL{%
\tempurl}


\bibitem[\protect\citeauthoryear{Broström}{Broström}{2022}]%
        {brostrom:yolo-strongsort}
\bibfield{author}{\bibinfo{person}{Mikel Broström}.}
  \bibinfo{year}{2022}\natexlab{}.
\newblock \bibinfo{title}{Real-time multi-camera multi-object tracker using
  YOLOv5 and StrongSORT with OSNet}.
\newblock
  \bibinfo{howpublished}{\url{https://github.com/mikel-brostrom/Yolov5_StrongSORT_OSNet}}.
\newblock


\bibitem[\protect\citeauthoryear{Caesar, Bankiti, Lang, Vora, Liong, Xu,
  Krishnan, Pan, Baldan, and Beijbom}{Caesar et~al\mbox{.}}{2019}]%
        {caesar:nuscenes}
\bibfield{author}{\bibinfo{person}{Holger Caesar}, \bibinfo{person}{Varun
  Bankiti}, \bibinfo{person}{Alex~H. Lang}, \bibinfo{person}{Sourabh Vora},
  \bibinfo{person}{Venice~Erin Liong}, \bibinfo{person}{Qiang Xu},
  \bibinfo{person}{Anush Krishnan}, \bibinfo{person}{Yu Pan},
  \bibinfo{person}{Giancarlo Baldan}, {and} \bibinfo{person}{Oscar Beijbom}.}
  \bibinfo{year}{2019}\natexlab{}.
\newblock \bibinfo{title}{nuScenes: A multimodal dataset for autonomous
  driving}.
\newblock
\newblock
\urldef\tempurl%
\url{https://doi.org/10.48550/ARXIV.1903.11027}
\showDOI{\tempurl}


\bibitem[\protect\citeauthoryear{Du, Zhao, Song, Zhao, Su, Gong, and Meng}{Du
  et~al\mbox{.}}{2023}]%
        {du:strongsort}
\bibfield{author}{\bibinfo{person}{Yunhao Du}, \bibinfo{person}{Zhicheng Zhao},
  \bibinfo{person}{Yang Song}, \bibinfo{person}{Yanyun Zhao},
  \bibinfo{person}{Fei Su}, \bibinfo{person}{Tao Gong}, {and}
  \bibinfo{person}{Hongying Meng}.} \bibinfo{year}{2023}\natexlab{}.
\newblock \bibinfo{title}{StrongSORT: Make DeepSORT Great Again}.
\newblock
\newblock
\showeprint[arxiv]{2202.13514}~[cs.CV]


\bibitem[\protect\citeauthoryear{Fremont, Dreossi, Ghosh, Yue,
  Sangiovanni-Vincentelli, and Seshia}{Fremont et~al\mbox{.}}{2019}]%
        {fremont:scenic-scene}
\bibfield{author}{\bibinfo{person}{Daniel~J. Fremont}, \bibinfo{person}{Tommaso
  Dreossi}, \bibinfo{person}{Shromona Ghosh}, \bibinfo{person}{Xiangyu Yue},
  \bibinfo{person}{Alberto~L. Sangiovanni-Vincentelli}, {and}
  \bibinfo{person}{Sanjit~A. Seshia}.} \bibinfo{year}{2019}\natexlab{}.
\newblock \showarticletitle{Scenic: A Language for Scenario Specification and
  Scene Generation}. In \bibinfo{booktitle}{\emph{Proceedings of the 40th ACM
  SIGPLAN Conference on Programming Language Design and Implementation}}
  (Phoenix, AZ, USA) \emph{(\bibinfo{series}{PLDI 2019})}.
  \bibinfo{publisher}{Association for Computing Machinery},
  \bibinfo{address}{New York, NY, USA}, \bibinfo{pages}{63–78}.
\newblock
\showISBNx{9781450367127}
\urldef\tempurl%
\url{https://doi.org/10.1145/3314221.3314633}
\showDOI{\tempurl}


\bibitem[\protect\citeauthoryear{Fremont, Kim, Dreossi, Ghosh, Yue,
  Sangiovanni-Vincentelli, and Seshia}{Fremont et~al\mbox{.}}{2023}]%
        {fremont:scenic-data}
\bibfield{author}{\bibinfo{person}{Daniel~J. Fremont}, \bibinfo{person}{Edward
  Kim}, \bibinfo{person}{Tommaso Dreossi}, \bibinfo{person}{Shromona Ghosh},
  \bibinfo{person}{Xiangyu Yue}, \bibinfo{person}{Alberto~L.
  Sangiovanni-Vincentelli}, {and} \bibinfo{person}{Sanjit~A. Seshia}.}
  \bibinfo{year}{2023}\natexlab{}.
\newblock \showarticletitle{Scenic: a language for scenario specification and
  data generation}.
\newblock \bibinfo{journal}{\emph{Machine Learning}} \bibinfo{volume}{112},
  \bibinfo{number}{10} (\bibinfo{date}{01 Oct} \bibinfo{year}{2023}),
  \bibinfo{pages}{3805--3849}.
\newblock
\showISSN{1573-0565}
\urldef\tempurl%
\url{https://doi.org/10.1007/s10994-021-06120-5}
\showDOI{\tempurl}


\bibitem[\protect\citeauthoryear{Fritsch, Kühnl, and Geiger}{Fritsch
  et~al\mbox{.}}{2013}]%
        {fritsch:kitti-road}
\bibfield{author}{\bibinfo{person}{Jannik Fritsch}, \bibinfo{person}{Tobias
  Kühnl}, {and} \bibinfo{person}{Andreas Geiger}.}
  \bibinfo{year}{2013}\natexlab{}.
\newblock \showarticletitle{A new performance measure and evaluation benchmark
  for road detection algorithms}. In \bibinfo{booktitle}{\emph{16th
  International IEEE Conference on Intelligent Transportation Systems (ITSC
  2013)}}. \bibinfo{pages}{1693--1700}.
\newblock
\showISSN{2153-0017}
\urldef\tempurl%
\url{https://doi.org/10.1109/ITSC.2013.6728473}
\showDOI{\tempurl}


\bibitem[\protect\citeauthoryear{Ge, Lin, Daum, Haynes, Cheung, and
  Balazinska}{Ge et~al\mbox{.}}{2021}]%
        {ge:apperception}
\bibfield{author}{\bibinfo{person}{Yongming Ge}, \bibinfo{person}{Vanessa Lin},
  \bibinfo{person}{Maureen Daum}, \bibinfo{person}{Brandon Haynes},
  \bibinfo{person}{Alvin Cheung}, {and} \bibinfo{person}{Magdalena
  Balazinska}.} \bibinfo{year}{2021}\natexlab{}.
\newblock \showarticletitle{Demonstration of Apperception: A Database
  Management System for Geospatial Video Data}.
\newblock \bibinfo{journal}{\emph{Proc. VLDB Endow.}} \bibinfo{volume}{14},
  \bibinfo{number}{12} (\bibinfo{date}{oct} \bibinfo{year}{2021}),
  \bibinfo{pages}{2767–2770}.
\newblock
\showISSN{2150-8097}
\urldef\tempurl%
\url{https://doi.org/10.14778/3476311.3476340}
\showDOI{\tempurl}


\bibitem[\protect\citeauthoryear{Girshick}{Girshick}{2015}]%
        {girshick:fast-rcnn}
\bibfield{author}{\bibinfo{person}{Ross Girshick}.}
  \bibinfo{year}{2015}\natexlab{}.
\newblock \bibinfo{title}{Fast R-CNN}.
\newblock
\newblock
\showeprint[arxiv]{1504.08083}~[cs.CV]


\bibitem[\protect\citeauthoryear{Girshick, Donahue, Darrell, and
  Malik}{Girshick et~al\mbox{.}}{2014}]%
        {girshick:rcnn}
\bibfield{author}{\bibinfo{person}{Ross Girshick}, \bibinfo{person}{Jeff
  Donahue}, \bibinfo{person}{Trevor Darrell}, {and} \bibinfo{person}{Jitendra
  Malik}.} \bibinfo{year}{2014}\natexlab{}.
\newblock \bibinfo{title}{Rich feature hierarchies for accurate object
  detection and semantic segmentation}.
\newblock
\newblock
\showeprint[arxiv]{1311.2524}~[cs.CV]


\bibitem[\protect\citeauthoryear{Gloudemans and Work}{Gloudemans and
  Work}{2021}]%
        {gloudemans:vehicle-turn-count}
\bibfield{author}{\bibinfo{person}{Derek Gloudemans} {and}
  \bibinfo{person}{Daniel~B. Work}.} \bibinfo{year}{2021}\natexlab{}.
\newblock \showarticletitle{Fast Vehicle Turning-Movement Counting using
  Localization-based Tracking}. In \bibinfo{booktitle}{\emph{2021 IEEE/CVF
  Conference on Computer Vision and Pattern Recognition Workshops (CVPRW)}}.
  \bibinfo{pages}{4150--4159}.
\newblock
\showISSN{2160-7516}
\urldef\tempurl%
\url{https://doi.org/10.1109/CVPRW53098.2021.00469}
\showDOI{\tempurl}


\bibitem[\protect\citeauthoryear{Godard, Aodha, Firman, and Brostow}{Godard
  et~al\mbox{.}}{2019}]%
        {godard:monodepth}
\bibfield{author}{\bibinfo{person}{Clément Godard}, \bibinfo{person}{Oisin~Mac
  Aodha}, \bibinfo{person}{Michael Firman}, {and} \bibinfo{person}{Gabriel
  Brostow}.} \bibinfo{year}{2019}\natexlab{}.
\newblock \bibinfo{title}{Digging Into Self-Supervised Monocular Depth
  Estimation}.
\newblock
\newblock
\showeprint[arxiv]{1806.01260}~[cs.CV]


\bibitem[\protect\citeauthoryear{Griffin and Miller}{Griffin and
  Miller}{2008}]%
        {griffin:amber}
\bibfield{author}{\bibinfo{person}{Timothy Griffin} {and}
  \bibinfo{person}{Monica~K. Miller}.} \bibinfo{year}{2008}\natexlab{}.
\newblock \showarticletitle{Child Abduction, AMBER Alert, and Crime Control
  Theater}.
\newblock \bibinfo{journal}{\emph{Criminal Justice Review}}
  \bibinfo{volume}{33}, \bibinfo{number}{2} (\bibinfo{year}{2008}),
  \bibinfo{pages}{159--176}.
\newblock
\urldef\tempurl%
\url{https://doi.org/10.1177/0734016808316778}
\showDOI{\tempurl}
\showeprint{https://doi.org/10.1177/0734016808316778}


\bibitem[\protect\citeauthoryear{Haynes, Daum, Mazumdar, Balazinska, Cheung,
  and Ceze}{Haynes et~al\mbox{.}}{2020}]%
        {haynes:visualworlddb}
\bibfield{author}{\bibinfo{person}{Brandon Haynes}, \bibinfo{person}{Maureen
  Daum}, \bibinfo{person}{Amrita Mazumdar}, \bibinfo{person}{Magdalena
  Balazinska}, \bibinfo{person}{Alvin Cheung}, {and} \bibinfo{person}{Luis
  Ceze}.} \bibinfo{year}{2020}\natexlab{}.
\newblock \showarticletitle{VisualWorldDB: A DBMS for the Visual World}. In
  \bibinfo{booktitle}{\emph{Conference on Innovative Data Systems Research}}.
  6.
\newblock


\bibitem[\protect\citeauthoryear{Jocher, Chaurasia, Stoken, Borovec,
  NanoCode012, Kwon, Michael, TaoXie, Fang, imyhxy, Lorna, Yifu), Wong, V,
  Montes, Wang, Fati, Nadar, Laughing, UnglvKitDe, Sonck, tkianai, yxNONG,
  Skalski, Hogan, Nair, Strobel, and Jain}{Jocher et~al\mbox{.}}{2022}]%
        {josher:yolov5}
\bibfield{author}{\bibinfo{person}{Glenn Jocher}, \bibinfo{person}{Ayush
  Chaurasia}, \bibinfo{person}{Alex Stoken}, \bibinfo{person}{Jirka Borovec},
  \bibinfo{person}{NanoCode012}, \bibinfo{person}{Yonghye Kwon},
  \bibinfo{person}{Kalen Michael}, \bibinfo{person}{TaoXie},
  \bibinfo{person}{Jiacong Fang}, \bibinfo{person}{imyhxy},
  \bibinfo{person}{Lorna}, \bibinfo{person}{曾逸夫(Zeng Yifu)},
  \bibinfo{person}{Colin Wong}, \bibinfo{person}{Abhiram V},
  \bibinfo{person}{Diego Montes}, \bibinfo{person}{Zhiqiang Wang},
  \bibinfo{person}{Cristi Fati}, \bibinfo{person}{Jebastin Nadar},
  \bibinfo{person}{Laughing}, \bibinfo{person}{UnglvKitDe},
  \bibinfo{person}{Victor Sonck}, \bibinfo{person}{tkianai},
  \bibinfo{person}{yxNONG}, \bibinfo{person}{Piotr Skalski},
  \bibinfo{person}{Adam Hogan}, \bibinfo{person}{Dhruv Nair},
  \bibinfo{person}{Max Strobel}, {and} \bibinfo{person}{Mrinal Jain}.}
  \bibinfo{year}{2022}\natexlab{}.
\newblock \bibinfo{booktitle}{\emph{{ultralytics/yolov5: v7.0 - YOLOv5 SOTA
  Realtime Instance Segmentation}}}.
\newblock
\urldef\tempurl%
\url{https://doi.org/10.5281/zenodo.7347926}
\showDOI{\tempurl}


\bibitem[\protect\citeauthoryear{Kang, Emmons, Abuzaid, Bailis, and
  Zaharia}{Kang et~al\mbox{.}}{2017}]%
        {kang:noscope}
\bibfield{author}{\bibinfo{person}{Daniel Kang}, \bibinfo{person}{John Emmons},
  \bibinfo{person}{Firas Abuzaid}, \bibinfo{person}{Peter Bailis}, {and}
  \bibinfo{person}{Matei Zaharia}.} \bibinfo{year}{2017}\natexlab{}.
\newblock \showarticletitle{NoScope: Optimizing Neural Network Queries over
  Video at Scale}.
\newblock \bibinfo{journal}{\emph{Proc. VLDB Endow.}} \bibinfo{volume}{10},
  \bibinfo{number}{11} (\bibinfo{date}{aug} \bibinfo{year}{2017}),
  \bibinfo{pages}{1586–1597}.
\newblock
\showISSN{2150-8097}
\urldef\tempurl%
\url{https://doi.org/10.14778/3137628.3137664}
\showDOI{\tempurl}


\bibitem[\protect\citeauthoryear{Kang, Romero, Bailis, Kozyrakis, and
  Zaharia}{Kang et~al\mbox{.}}{2022}]%
        {kang:viva}
\bibfield{author}{\bibinfo{person}{Daniel Kang}, \bibinfo{person}{Francisco
  Romero}, \bibinfo{person}{Peter~D. Bailis}, \bibinfo{person}{Christos
  Kozyrakis}, {and} \bibinfo{person}{Matei Zaharia}.}
  \bibinfo{year}{2022}\natexlab{}.
\newblock \showarticletitle{{VIVA:} An End-to-End System for Interactive Video
  Analytics}. In \bibinfo{booktitle}{\emph{12th Conference on Innovative Data
  Systems Research, {CIDR} 2022, Chaminade, CA, USA, January 9-12, 2022}}.
  \bibinfo{publisher}{www.cidrdb.org}, 9.
\newblock
\urldef\tempurl%
\url{https://www.cidrdb.org/cidr2022/papers/p75-kang.pdf}
\showURL{%
\tempurl}


\bibitem[\protect\citeauthoryear{Kim, Ošep, and Leal-Taixé}{Kim
  et~al\mbox{.}}{2021a}]%
        {kim:eagermot}
\bibfield{author}{\bibinfo{person}{Aleksandr Kim}, \bibinfo{person}{Aljoša
  Ošep}, {and} \bibinfo{person}{Laura Leal-Taixé}.}
  \bibinfo{year}{2021}\natexlab{a}.
\newblock \bibinfo{title}{EagerMOT: 3D Multi-Object Tracking via Sensor
  Fusion}.
\newblock
\newblock
\showeprint[arxiv]{2104.14682}~[cs.CV]


\bibitem[\protect\citeauthoryear{Kim, Shenoy, Junges, Fremont,
  Sangiovanni-Vincentelli, and Seshia}{Kim et~al\mbox{.}}{2021b}]%
        {kim:scenic-val}
\bibfield{author}{\bibinfo{person}{Edward Kim}, \bibinfo{person}{Jay Shenoy},
  \bibinfo{person}{Sebastian Junges}, \bibinfo{person}{Daniel Fremont},
  \bibinfo{person}{Alberto Sangiovanni-Vincentelli}, {and}
  \bibinfo{person}{Sanjit Seshia}.} \bibinfo{year}{2021}\natexlab{b}.
\newblock \bibinfo{title}{Querying Labelled Data with Scenario Programs for
  Sim-to-Real Validation}.
\newblock
\newblock
\showeprint[arxiv]{2112.00206}~[cs.CV]


\bibitem[\protect\citeauthoryear{Kuhn}{Kuhn}{1955}]%
        {kuhn:hungarian}
\bibfield{author}{\bibinfo{person}{H.~W. Kuhn}.}
  \bibinfo{year}{1955}\natexlab{}.
\newblock \showarticletitle{The Hungarian method for the assignment problem}.
\newblock \bibinfo{journal}{\emph{Naval Research Logistics Quarterly}}
  \bibinfo{volume}{2}, \bibinfo{number}{1-2} (\bibinfo{year}{1955}),
  \bibinfo{pages}{83--97}.
\newblock
\urldef\tempurl%
\url{https://doi.org/10.1002/nav.3800020109}
\showDOI{\tempurl}
\showeprint{https://onlinelibrary.wiley.com/doi/pdf/10.1002/nav.3800020109}


\bibitem[\protect\citeauthoryear{Kuipers}{Kuipers}{1999}]%
        {kuipers:quaternion}
\bibfield{author}{\bibinfo{person}{Jack~B. Kuipers}.}
  \bibinfo{year}{1999}\natexlab{}.
\newblock \bibinfo{booktitle}{\emph{Quaternions and rotation sequences : a
  primer with applications to orbits, aerospace, and virtual reality}}.
\newblock \bibinfo{publisher}{Princeton Univ. Press},
  \bibinfo{address}{Princeton, NJ}.
\newblock
\showISBNx{0691058725 9780691058726}
\urldef\tempurl%
\url{http://www.worldcat.org/title/quaternions-and-rotation-sequences-a-primer-with-applications-to-orbits-aerospace-and-virtual-reality/oclc/246446345}
\showURL{%
\tempurl}


\bibitem[\protect\citeauthoryear{Leal-Taixé}{Leal-Taixé}{2014}]%
        {leal-taixe:track-by-detect}
\bibfield{author}{\bibinfo{person}{Laura Leal-Taixé}.}
  \bibinfo{year}{2014}\natexlab{}.
\newblock \showarticletitle{Multiple object tracking with context awareness}.
\newblock  (\bibinfo{date}{11} \bibinfo{year}{2014}), \bibinfo{pages}{15--25}.
\newblock


\bibitem[\protect\citeauthoryear{Luiten, O\u{s}ep, Dendorfer, Torr, Geiger,
  Leal-Taix{\'{e}}, and Leibe}{Luiten et~al\mbox{.}}{2020}]%
        {luiten:hota}
\bibfield{author}{\bibinfo{person}{Jonathon Luiten},
  \bibinfo{person}{Aljo\u{s}a O\u{s}ep}, \bibinfo{person}{Patrick Dendorfer},
  \bibinfo{person}{Philip Torr}, \bibinfo{person}{Andreas Geiger},
  \bibinfo{person}{Laura Leal-Taix{\'{e}}}, {and} \bibinfo{person}{Bastian
  Leibe}.} \bibinfo{year}{2020}\natexlab{}.
\newblock \showarticletitle{{HOTA}: A Higher Order Metric for Evaluating
  Multi-object Tracking}.
\newblock \bibinfo{journal}{\emph{International Journal of Computer Vision}}
  \bibinfo{volume}{129}, \bibinfo{number}{2} (\bibinfo{date}{oct}
  \bibinfo{year}{2020}), \bibinfo{pages}{548--578}.
\newblock
\urldef\tempurl%
\url{https://doi.org/10.1007/s11263-020-01375-2}
\showDOI{\tempurl}


\bibitem[\protect\citeauthoryear{McDonald, Costello, Bone, Cabral, Farabee,
  Hochberg, Kroodsma, Mangin, Meng, and Zahn}{McDonald et~al\mbox{.}}{2021}]%
        {mcdonald:satellites-fishing}
\bibfield{author}{\bibinfo{person}{Gavin~G. McDonald},
  \bibinfo{person}{Christopher Costello}, \bibinfo{person}{Jennifer Bone},
  \bibinfo{person}{Reniel~B. Cabral}, \bibinfo{person}{Valerie Farabee},
  \bibinfo{person}{Timothy Hochberg}, \bibinfo{person}{David Kroodsma},
  \bibinfo{person}{Tracey Mangin}, \bibinfo{person}{Kyle~C. Meng}, {and}
  \bibinfo{person}{Oliver Zahn}.} \bibinfo{year}{2021}\natexlab{}.
\newblock \showarticletitle{Satellites can reveal global extent of forced labor
  in the world’s fishing fleet}.
\newblock \bibinfo{journal}{\emph{Proceedings of the National Academy of
  Sciences}} \bibinfo{volume}{118}, \bibinfo{number}{3} (\bibinfo{year}{2021}),
  \bibinfo{pages}{e2016238117}.
\newblock
\urldef\tempurl%
\url{https://doi.org/10.1073/pnas.2016238117}
\showDOI{\tempurl}
\showeprint{https://www.pnas.org/doi/pdf/10.1073/pnas.2016238117}


\bibitem[\protect\citeauthoryear{Park, Lee, Seto, Hochberg, Wong, Miller,
  Takasaki, Kubota, Oozeki, Doshi, Midzik, Hanich, Sullivan, Woods, and
  Kroodsma}{Park et~al\mbox{.}}{2020}]%
        {park:satellite-fishing}
\bibfield{author}{\bibinfo{person}{Jaeyoon Park}, \bibinfo{person}{Jungsam
  Lee}, \bibinfo{person}{Katherine Seto}, \bibinfo{person}{Timothy Hochberg},
  \bibinfo{person}{Brian~A. Wong}, \bibinfo{person}{Nathan~A. Miller},
  \bibinfo{person}{Kenji Takasaki}, \bibinfo{person}{Hiroshi Kubota},
  \bibinfo{person}{Yoshioki Oozeki}, \bibinfo{person}{Sejal Doshi},
  \bibinfo{person}{Maya Midzik}, \bibinfo{person}{Quentin Hanich},
  \bibinfo{person}{Brian Sullivan}, \bibinfo{person}{Paul Woods}, {and}
  \bibinfo{person}{David~A. Kroodsma}.} \bibinfo{year}{2020}\natexlab{}.
\newblock \showarticletitle{Illuminating dark fishing fleets in North Korea}.
\newblock \bibinfo{journal}{\emph{Science Advances}} \bibinfo{volume}{6},
  \bibinfo{number}{30} (\bibinfo{year}{2020}), \bibinfo{pages}{eabb1197}.
\newblock
\urldef\tempurl%
\url{https://doi.org/10.1126/sciadv.abb1197}
\showDOI{\tempurl}
\showeprint{https://www.science.org/doi/pdf/10.1126/sciadv.abb1197}


\bibitem[\protect\citeauthoryear{Redmon and Farhadi}{Redmon and
  Farhadi}{2018}]%
        {redmon:yolov3}
\bibfield{author}{\bibinfo{person}{Joseph Redmon} {and} \bibinfo{person}{Ali
  Farhadi}.} \bibinfo{year}{2018}\natexlab{}.
\newblock \bibinfo{title}{YOLOv3: An Incremental Improvement}.
\newblock
\newblock
\showeprint[arxiv]{1804.02767}~[cs.CV]


\bibitem[\protect\citeauthoryear{Ren, He, Girshick, and Sun}{Ren
  et~al\mbox{.}}{2016}]%
        {ren:faster-rcnn}
\bibfield{author}{\bibinfo{person}{Shaoqing Ren}, \bibinfo{person}{Kaiming He},
  \bibinfo{person}{Ross Girshick}, {and} \bibinfo{person}{Jian Sun}.}
  \bibinfo{year}{2016}\natexlab{}.
\newblock \bibinfo{title}{Faster R-CNN: Towards Real-Time Object Detection with
  Region Proposal Networks}.
\newblock
\newblock
\showeprint[arxiv]{1506.01497}~[cs.CV]


\bibitem[\protect\citeauthoryear{Romero, Hauswald, Partap, Kang, Zaharia, and
  Kozyrakis}{Romero et~al\mbox{.}}{2022}]%
        {romero:viva}
\bibfield{author}{\bibinfo{person}{Francisco Romero}, \bibinfo{person}{Johann
  Hauswald}, \bibinfo{person}{Aditi Partap}, \bibinfo{person}{Daniel Kang},
  \bibinfo{person}{Matei Zaharia}, {and} \bibinfo{person}{Christos Kozyrakis}.}
  \bibinfo{year}{2022}\natexlab{}.
\newblock \showarticletitle{Optimizing Video Analytics with Declarative Model
  Relationships}.
\newblock \bibinfo{journal}{\emph{Proc. VLDB Endow.}} \bibinfo{volume}{16},
  \bibinfo{number}{3} (\bibinfo{date}{nov} \bibinfo{year}{2022}),
  \bibinfo{pages}{447–460}.
\newblock
\showISSN{2150-8097}
\urldef\tempurl%
\url{https://doi.org/10.14778/3570690.3570695}
\showDOI{\tempurl}


\bibitem[\protect\citeauthoryear{Simon, Amende, Kraus, Honer, Sämann,
  Kaulbersch, Milz, and Gross}{Simon et~al\mbox{.}}{2019}]%
        {simon:complexer-yolo}
\bibfield{author}{\bibinfo{person}{Martin Simon}, \bibinfo{person}{Karl
  Amende}, \bibinfo{person}{Andrea Kraus}, \bibinfo{person}{Jens Honer},
  \bibinfo{person}{Timo Sämann}, \bibinfo{person}{Hauke Kaulbersch},
  \bibinfo{person}{Stefan Milz}, {and} \bibinfo{person}{Horst~Michael Gross}.}
  \bibinfo{year}{2019}\natexlab{}.
\newblock \bibinfo{title}{Complexer-YOLO: Real-Time 3D Object Detection and
  Tracking on Semantic Point Clouds}.
\newblock
\newblock
\showeprint[arxiv]{1904.07537}~[cs.CV]


\bibitem[\protect\citeauthoryear{Sklansky}{Sklansky}{1982}]%
        {sklansky:convexhull}
\bibfield{author}{\bibinfo{person}{Jack Sklansky}.}
  \bibinfo{year}{1982}\natexlab{}.
\newblock \showarticletitle{Finding the Convex Hull of a Simple Polygon}.
\newblock \bibinfo{journal}{\emph{Pattern Recogn. Lett.}} \bibinfo{volume}{1},
  \bibinfo{number}{2} (\bibinfo{date}{dec} \bibinfo{year}{1982}),
  \bibinfo{pages}{79–83}.
\newblock
\showISSN{0167-8655}
\urldef\tempurl%
\url{https://doi.org/10.1016/0167-8655(82)90016-2}
\showDOI{\tempurl}


\bibitem[\protect\citeauthoryear{Sparks}{Sparks}{2023}]%
        {sparks:security-cam}
\bibfield{author}{\bibinfo{person}{Robert Sparks}.}
  \bibinfo{year}{2023}\natexlab{}.
\newblock \bibinfo{title}{11 Security Camera Statistics \& Data - 2023 update}.
\newblock
\newblock
\urldef\tempurl%
\url{https://opticsmag.com/security-camera-statistics/}
\showURL{%
\tempurl}


\bibitem[\protect\citeauthoryear{Tomar}{Tomar}{2006}]%
        {suramya:ffmpeg}
\bibfield{author}{\bibinfo{person}{Suramya Tomar}.}
  \bibinfo{year}{2006}\natexlab{}.
\newblock \showarticletitle{Converting Video Formats with FFmpeg}.
\newblock \bibinfo{journal}{\emph{Linux J.}} \bibinfo{volume}{2006},
  \bibinfo{number}{146} (\bibinfo{date}{jun} \bibinfo{year}{2006}),
  \bibinfo{pages}{10}.
\newblock
\showISSN{1075-3583}


\bibitem[\protect\citeauthoryear{Wojke, Bewley, and Paulus}{Wojke
  et~al\mbox{.}}{2017}]%
        {wojke:deepsort}
\bibfield{author}{\bibinfo{person}{Nicolai Wojke}, \bibinfo{person}{Alex
  Bewley}, {and} \bibinfo{person}{Dietrich Paulus}.}
  \bibinfo{year}{2017}\natexlab{}.
\newblock \bibinfo{title}{Simple Online and Realtime Tracking with a Deep
  Association Metric}.
\newblock
\newblock
\showeprint[arxiv]{1703.07402}~[cs.CV]


\bibitem[\protect\citeauthoryear{Wright}{Wright}{2023}]%
        {wright:av-data}
\bibfield{author}{\bibinfo{person}{Simon Wright}.}
  \bibinfo{year}{2023}\natexlab{}.
\newblock \bibinfo{title}{Autonomous cars generate more than 300 TB of data per
  year}.
\newblock
\newblock
\urldef\tempurl%
\url{https://www.tuxera.com/blog/autonomous-cars-300-tb-of-data-per-year/}
\showURL{%
\tempurl}


\bibitem[\protect\citeauthoryear{Xu, Kakkar, Arulraj, and Ramachandran}{Xu
  et~al\mbox{.}}{2022}]%
        {xu:eva}
\bibfield{author}{\bibinfo{person}{Zhuangdi Xu}, \bibinfo{person}{Gaurav~Tarlok
  Kakkar}, \bibinfo{person}{Joy Arulraj}, {and} \bibinfo{person}{Umakishore
  Ramachandran}.} \bibinfo{year}{2022}\natexlab{}.
\newblock \showarticletitle{EVA: A Symbolic Approach to Accelerating
  Exploratory Video Analytics with Materialized Views}. In
  \bibinfo{booktitle}{\emph{Proceedings of the 2022 International Conference on
  Management of Data}} (Philadelphia, PA, USA) \emph{(\bibinfo{series}{SIGMOD
  '22})}. \bibinfo{publisher}{Association for Computing Machinery},
  \bibinfo{address}{New York, NY, USA}, \bibinfo{pages}{602–616}.
\newblock
\showISBNx{9781450392495}
\urldef\tempurl%
\url{https://doi.org/10.1145/3514221.3526142}
\showDOI{\tempurl}


\bibitem[\protect\citeauthoryear{Zhang}{Zhang}{2014}]%
        {zhang:camera}
\bibfield{author}{\bibinfo{person}{Zhengyou Zhang}.}
  \bibinfo{year}{2014}\natexlab{}.
\newblock \bibinfo{booktitle}{\emph{Camera Parameters (Intrinsic, Extrinsic)}}.
\newblock \bibinfo{publisher}{Springer US}, \bibinfo{address}{Boston, MA},
  \bibinfo{pages}{81--85}.
\newblock
\showISBNx{978-0-387-31439-6}
\urldef\tempurl%
\url{https://doi.org/10.1007/978-0-387-31439-6_152}
\showDOI{\tempurl}


\bibitem[\protect\citeauthoryear{Zhou, Yang, Cavallaro, and Xiang}{Zhou
  et~al\mbox{.}}{2019}]%
        {zhou:osnet}
\bibfield{author}{\bibinfo{person}{Kaiyang Zhou}, \bibinfo{person}{Yongxin
  Yang}, \bibinfo{person}{Andrea Cavallaro}, {and} \bibinfo{person}{Tao
  Xiang}.} \bibinfo{year}{2019}\natexlab{}.
\newblock \bibinfo{title}{Omni-Scale Feature Learning for Person
  Re-Identification}.
\newblock
\newblock
\showeprint[arxiv]{1905.00953}~[cs.CV]


\bibitem[\protect\citeauthoryear{Zim\'{a}nyi, Sakr, and Lesuisse}{Zim\'{a}nyi
  et~al\mbox{.}}{2020}]%
        {zimanyi:mobility}
\bibfield{author}{\bibinfo{person}{Esteban Zim\'{a}nyi},
  \bibinfo{person}{Mahmoud Sakr}, {and} \bibinfo{person}{Arthur Lesuisse}.}
  \bibinfo{year}{2020}\natexlab{}.
\newblock \showarticletitle{MobilityDB: A Mobility Database Based on PostgreSQL
  and PostGIS}.
\newblock \bibinfo{journal}{\emph{ACM Trans. Database Syst.}}
  \bibinfo{volume}{45}, \bibinfo{number}{4}, Article \bibinfo{articleno}{19}
  (\bibinfo{date}{dec} \bibinfo{year}{2020}), \bibinfo{numpages}{42}~pages.
\newblock
\showISSN{0362-5915}
\urldef\tempurl%
\url{https://doi.org/10.1145/3406534}
\showDOI{\tempurl}


\end{thebibliography}
\end{CJK*}
\end{document}